\documentclass[prl,twocolumn,amsmath,amssymb,superscriptaddress,showpacs]{revtex4-2}
\usepackage[english]{babel}
\usepackage[T1]{fontenc} 
\usepackage[hypertexnames=false]{hyperref}
\usepackage{graphicx}
\usepackage{bbm}
\usepackage{orcidlink}

\usepackage{bm} 

\newcommand{\cphantom}[2]{
  \sbox0{#2}                
  \makebox[\wd0][c]{#1}     
}

\makeatletter
\newcommand{\textlabel}[2]{%
  \def\@currentlabel{#2}
  \label{#1}
}
\makeatother

\newcounter{myequation}
\makeatletter
\@addtoreset{equation}{myequation}
\makeatother

\newcounter{myfigure}
\makeatletter
\@addtoreset{figure}{myfigure}
\makeatother

\newcounter{mytable}
\makeatletter
\@addtoreset{table}{mytable}
\makeatother

\newcounter{mytheorem}
\makeatletter
\@addtoreset{theorem}{mytheorem}
\makeatother

\newcounter{mysection}
\makeatletter
\@addtoreset{section}{mysection}
\makeatother

\newcommand{\ket}[1]{|#1\rangle}
\newcommand{\ketbra}[1]{| #1\rangle \langle #1|}

\newcommand{\be}{\begin{equation}}
\newcommand{\ee}{\end{equation}}
\newcommand{\eea}{\end{eqnarray}}
\newcommand{\bea}{\begin{eqnarray}}

\newcommand{\va}[1]{\ensuremath{(\Delta#1)^2}}

\newcommand{\ex}[1]{\ensuremath{\langle{#1}\rangle}}

\newcommand{\qed}{\ensuremath{\hfill \blacksquare}}

\newcommand{\kommentar}[1]{}

\newcommand{\forget}[1]{}

\newcommand{\EQ}[1]{Eq.~\eqref{#1}}
\newcommand{\EQS}[1]{Eqs.~\eqref{#1}}
\newcommand{\EQL}[1]{Equation~\eqref{#1}}

\newcommand{\FIG}[1]{Fig.~\ref{#1}}
\newcommand{\REF}[1]{Ref.~\cite{#1}}

\newcommand{\THM}[1]{Theorem~\ref{#1}}

\newcommand{\DEFTHM}[1]{{\bf Theorem \refstepcounter{theorem}\thetheorem\label{#1}.} }
\newcounter{theorem}

\newcounter{lemma}

\newcounter{example}

\newcounter{definition}

\begin{document}

\title{Quantum Fisher Information and Separable Energy Minima in Spin Systems}
\author{G\'eza T\'oth\,\orcidlink{0000-0002-9602-751X}}
\email{toth@alumni.nd.edu}
\homepage{https://www.gtoth.eu}
\affiliation{Theoretical Physics, University of the Basque Country UPV/EHU,  
48080 Bilbao, Spain}
\affiliation{EHU Quantum Center, University of the Basque Country UPV/EHU, 
48940 Leioa, 
Spain}
\affiliation{Donostia International Physics Center DIPC,  
20018 San Sebasti\'an, Spain}
\affiliation{IKERBASQUE, Basque Foundation for Science, 48009 Bilbao, Spain}
\affiliation{HUN-REN Wigner Research Centre for Physics,  
1525 Budapest, Hungary}

\author{J\'ozsef Pitrik}
\email{pitrik@math.bme.hu}
\affiliation{Department of Analysis and Operations Research, Institute of Mathematics, Budapest University of Technology and Economics, 
1111 Budapest, Hungary}
\affiliation{HUN-REN 
R\'enyi Institute of Mathematics, 
1053 Budapest, Hungary}
\affiliation{HUN-REN Wigner Research Centre for Physics,  
1525 Budapest, Hungary}

\begin{abstract}
We develop a general method to bound the minimum energy of spin Hamiltonians over separable states with fixed single-particle reduced density matrices. The method uses a two-particle optimization and identifies conditions under which the resulting bound is tight. For ferromagnetic Ising and Ising-like interactions on any regular graph, we obtain an exact separable-energy minimum expressed through the quantum Fisher information (QFI) of the local state; a symmetric separable state attains the optimal correlation on every bond. For fully connected ferromagnetic Ising-like systems, ground-state correlations give a lower estimate of this single-particle QFI with a rigorous $O(1/N)$ finite-size error. Further exact results relate symmetric-separable minima of antiferromagnetic Ising-like and XY-like models to local expectation values, and ferromagnetic spin-$1/2$ Heisenberg chains to quantum fidelity. We also derive bounds for separability across fixed partitions into contiguous blocks.
\end{abstract}

\date{\today}

\maketitle

{\it Introduction.} The energy of a spin Hamiltonian can witness entanglement: any state with energy below the separable minimum must be entangled \cite{Brukner2004MacroscopicB,Toth2005EntanglementWitnesses,Toth2006Detection,Guhne2006Energy,Guhne2005Multipartite,Dowling2004Energy,Wu2005Entanglement,Igloi2023Entanglement}. This idea works especially well for many-body systems, but the corresponding energy thresholds have generally been derived on a case-by-case basis, without exploiting knowledge of the single-particle reduced density matrices. Incorporating those marginals turns the problem into a constrained many-body separability optimization.

Here we establish a connection between the energetics of spin systems and quantum information by relating the energy minima for separable states to the \emph{quantum Fisher information} (QFI) \cite{Helstrom1976Quantum,Holevo1982Probabilistic,Braunstein1994Statistical,Braunstein1996Generalized,Petz2008Quantum,Toth2014Quantum} of the single-particle states. Using the convex-roof representation of QFI and fixed-marginal separable-state optimization \cite{Toth2013Extremal,Yu2013Quantum,Toth2023QuantumWasserstein}, we reduce the constrained many-body problem to a two-particle optimization and identify when its optimum can be attained throughout the interaction graph. For ferromagnetic Ising and Ising-like interactions, a symmetric separable state attains the optimal correlation on every bond of any regular graph, yielding an exact many-body energy minimum.

Our approach also provides a route to estimating single-particle QFI from many-body correlations with a controlled finite-size error. In a fully connected ferromagnetic model, two-body correlations and the local second moment of a zero-temperature ground-state ensemble yield a QFI lower estimate with $O(1/N)$ error. We also obtain exact results in antiferromagnetic Ising-like and XY-like models and in ferromagnetic Heisenberg chains, where the Uhlmann--Jozsa fidelity enters \cite{Uhlmann1976TheTransitionProbability,Jozsa1994Fidelity}.

{\it Exact ferromagnetic energy minimum.} Consider $N$ particles of local dimension $d$ on a regular interaction graph,
\begin{equation}
H=\sum_{\langle n,n'\rangle}\sum_{l=1}^{L}J_l h_l^{(n)}h_l^{(n')}-\vec B\vec G,\label{eq:H}
\end{equation}
where $J_l$ are real couplings, $h_l^{(n)}$ acts on particle $n$, and $\vec G=\sum_n\vec g^{(n)}$ collects single-particle observables. We take $\vec g$ to be the $SU(d)$ generators; for qubits these may be the Pauli matrices. There are $N_p$ interacting pairs, so each site has $2N_p/N$ neighbors.

We prescribe the same marginal $\varrho$ at every site and denote the compatible separable states by $\mathrm{Sep}_N(\varrho)$, with minimum energy $E_{\rm sep}(\varrho)$. The smaller set $\mathrm{SymSep}_N(\varrho)$ consists of mixtures of identical pure products, $|\Psi\rangle\langle\Psi|^{\otimes N}$, with minimum energy $E_{\rm symsep}(\varrho)$. These states are supported on the symmetric subspace; permutation invariance alone does not imply this property.

\DEFTHM{thm:esepbound}For $L=1$ and $J_1=-J<0$ on any regular interaction graph,
\begin{equation}
\begin{aligned}
E_{\rm sep}(\varrho)&=E_{\rm symsep}(\varrho)\\
&=-JN_p\left(\ex{h_1^2}_\varrho-\frac14\mathcal F_Q[\varrho,h_1]\right)
-N\vec B\ex{\vec g}_\varrho.
\end{aligned}\label{eq:Esepvarrho}
\end{equation}
The result holds for arbitrary local dimension.

At fixed $\varrho$, the terms depending only on the marginal are already determined. Minimizing the ferromagnetic interaction therefore means making each bond correlation as large as separability allows. Equation~\eqref{eq:Esepvarrho} identifies that maximum: the local second moment $\ex{h_1^2}_\varrho$ is reduced by the minimum ensemble-averaged variance, $\mathcal F_Q[\varrho,h_1]/4$. 

Two limits make this restriction concrete. If $[\varrho,h_1]=0$, an ensemble of common eigenstates has zero QFI and attains the correlation $\ex{h_1^2}_\varrho$. If $\varrho$ is pure, the only compatible many-body state is $\varrho^{\otimes N}$, and the correlation is $\ex{h_1}_\varrho^2$. For a mixed marginal, the maximum separable correlation is attained by a pure-state decomposition minimizing the average variance of $h_1.$

{\it Proof.} The key step is the convex-roof representation of QFI \cite{Toth2013Extremal,Yu2013Quantum},
\begin{align*}
\mathcal F_Q[\varrho,h]
&=4\min_{\{p_k,\Psi_k\}}\sum_kp_k\va{h}_{\Psi_k}\\
&=4\min_{\omega_{12}}
\ex{h^2\otimes\openone-h\otimes h}_{\omega_{12}}\\
&=4\ex{h^2}_{\varrho}
-4\max_{\omega_{12}}\ex{h\otimes h}_{\omega_{12}}.
\end{align*}
The first minimum runs over pure-state decompositions $\varrho=\sum_kp_k\ketbra{\Psi_k}$. Each decomposition defines a symmetric separable state $\omega_{12}=\sum_kp_k\ketbra{\Psi_k}^{\otimes2}$ with both marginals $\varrho$. Its correlation is the ensemble average of $\ex{h}_{\Psi_k}^2$. Since the second moment is fixed, minimizing the average variance maximizes this correlation.

To compare with a general separable state, let us consider $\varrho_{AB}=\sum_kp_k\ketbra{\Psi_k}\otimes\ketbra{\Phi_k}$ and set $x_k=\ex{h}_{\Psi_k}$, $y_k=\ex{h}_{\Phi_k}$. Then we obtain
\[
\sum_kp_kx_ky_k\le\frac12\sum_kp_k(x_k^2+y_k^2)
\le\ex{h^2}_\varrho-\frac14\mathcal F_Q[\varrho,h].
\]
The first inequality holds term by term. The second inequality follows by applying the convex-roof formula for QFI separately to the two pure-state decompositions of $\varrho.$ Hence [Table~\ref{tab:max}(a)]
\begin{equation}
\begin{aligned}
\max_{\varrho_{AB}\in\mathrm{Sep}_2(\varrho)}\ex{h\otimes h}_{\varrho_{AB}}
&=\max_{\varrho_{AB}\in\mathrm{SymSep}_2(\varrho)}\ex{h\otimes h}_{\varrho_{AB}}\\
&=\ex{h^2}_\varrho-\frac14\mathcal F_Q[\varrho,h].
\end{aligned}\label{eq:maxhhFq}
\end{equation}
Every bond of a separable many-body state obeys this bound. An optimizing decomposition of $\varrho$ attains it simultaneously on every bond through
\[
\Omega_N=\sum_kp_k\ketbra{\Psi_k}^{\otimes N}.
\]
This state has the prescribed marginal at every site and belongs to both separable classes, proving Eq.~\eqref{eq:Esepvarrho}. \qed

The essential many-body step is the simultaneous saturation: the two-particle correlation identity alone would only bound each bond separately. Here a single globally consistent mixture realizes its maximum on \emph{all} bonds, independently of the regular graph's geometry. The result therefore supplies an exact entanglement-energy threshold, rather than merely an efficiently computable lower bound. For mixed $\varrho$, the saturating state can be a mixture of identical-site pure products rather than a single product. This state has the prescribed single-particle marginal $\varrho$ at every site and attains the maximum separable correlation on every bond. Thus the exact separable threshold depends on graph geometry only through $N_p$, even though the ground state of the Hamiltonian can depend strongly on the graph.

\begin{figure}[t!]
\includegraphics[width=0.45\columnwidth]{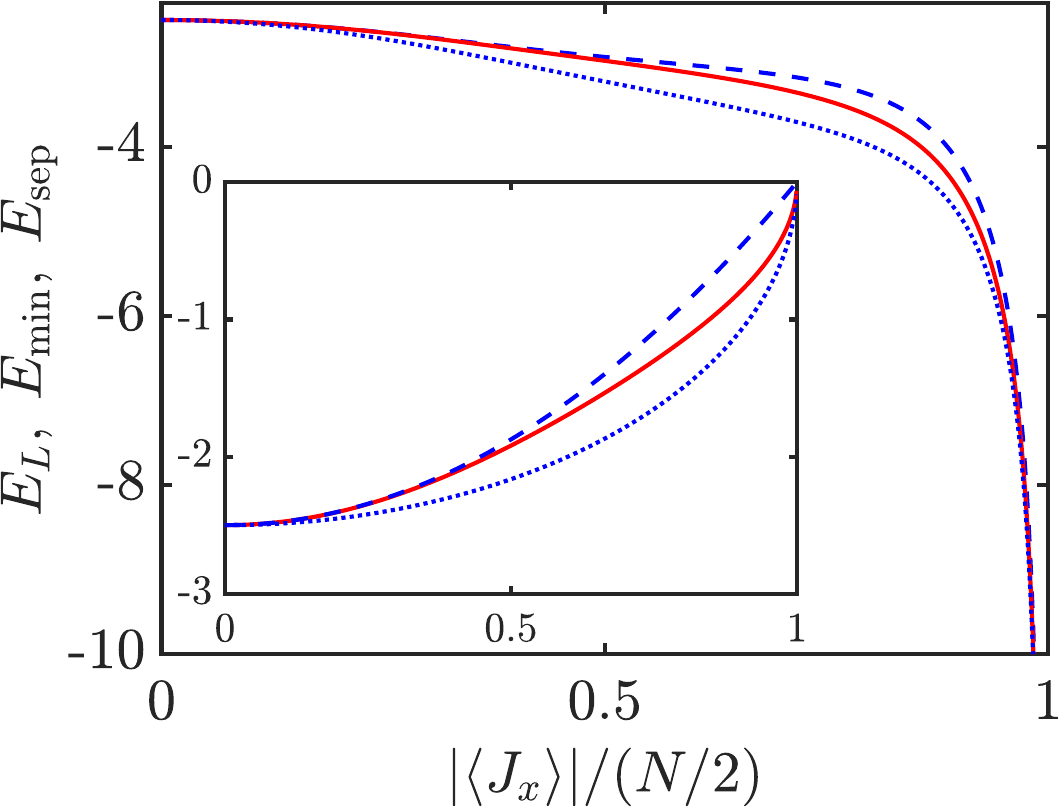}\hskip0.5cm
\includegraphics[width=0.45\columnwidth]{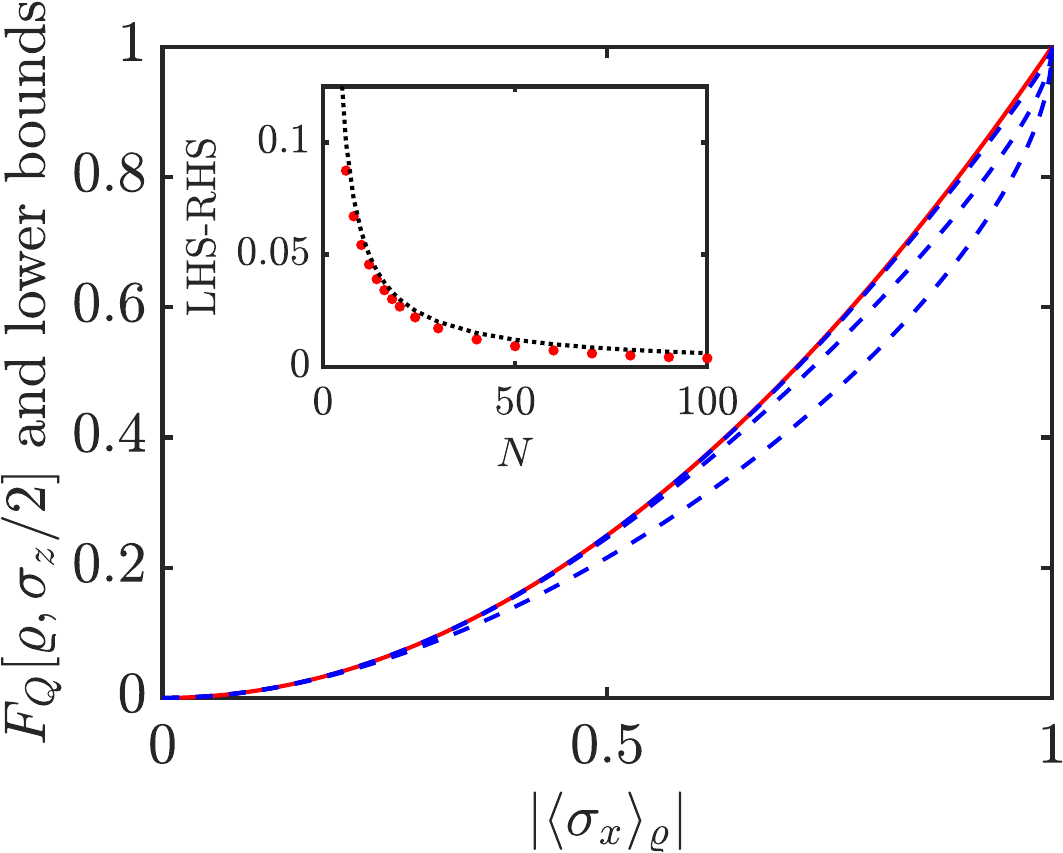}

\hspace*{0.50cm} (a) \hspace{3.7cm} (b)

\caption{(a) Energy bounds for the ferromagnetic Ising chain versus $\ex{J_x}$. Solid: ground-state energy; dashed: exact separable threshold at the ground-state marginal; dotted: lower bound for arbitrary states from Wigner--Yanase skew information. Inset: interaction energies without the field term. (b) Single-particle QFI $\mathcal F_Q[\varrho,\sigma_z/2]$ (solid) versus $|\ex{\sigma_x}_\varrho|$, with correlation-based lower estimates (dashed, bottom to top: $N=4,10,60$). Inset: estimation error (disks) versus $N$, with $0.6/N$ as a guide (dotted).}
\label{fig:FQ_ground}
\end{figure}

{\it Energy bounds and entanglement.} Figure~\ref{fig:FQ_ground}(a) illustrates the result for a periodic ferromagnetic Ising chain of $N=10$ spins with $J=1$ and $\vec B=(B_x,0,0)$. For each value of the magnetic field, we calculate the separable-energy threshold for the single-particle reduced density matrix of the corresponding ground state. The solid curve gives its energy and the dashed curve the exact separable threshold. 

The dotted curve provides a complementary lower bound $E_{\rm LB}(\varrho)$ obtained by allowing entangled two-particle states. For qubits, this bound follows from Eq.~\eqref{eq:Esepvarrho} by replacing $\mathcal F_Q/4$ with the Wigner--Yanase skew information \cite{Wigner1963INFORMATION}; see Table~\ref{tab:max}(c) and Supplemental Material~\ref{sec:entcond}. When the unconstrained ground state has marginal $\varrho$ at every site,
\begin{equation}
E_{\rm sep}(\varrho)\ge E_{\min}\ge E_{\rm LB}(\varrho).\label{eq:EsepEEL}
\end{equation}
The figure thus brackets the ground-state energy while comparing it with the separable threshold. Its inset removes the field contribution and displays the interaction energies alone.

Whenever the ground-state energy is below the separable-energy threshold, the ground state is detected as entangled. 
The same test applies to a thermal state $\varrho_{N,T}$ with marginal $\varrho_T$: entanglement is certified when $\ex{H}_{\varrho_{N,T}}<E_{\rm sep}(\varrho_T)$. In the separate $N=8$ example in Supplemental Material~\ref{sec:entcond}, Fig.~\ref{fig:FQ_temperature_bound} shows that local information extends the temperature range of detection. The improvement comes from excluding separable states with incompatible marginals.

{\it Quantum Fisher information from correlations.} The energy--QFI relation can also be used to estimate the QFI. Let us consider the ferromagnetic Ising-like model ($L=1$) on the complete graph, where $N_p=N(N-1)/2$. Let us prepare its zero-temperature thermal state $\omega_N$ and denote the resulting single-particle marginal by $\varrho$. If the ground level is degenerate, $\omega_N$ is the normalized projector onto that subspace. This choice is permutation invariant.

The measured quantity is the average bond correlation,
\[
C_N(\varrho)=\frac{1}{N(N-1)}\sum_{m\ne n}
\ex{h_1^{(m)}h_1^{(n)}}_{\omega_N}.
\]
At fixed marginal the field contribution is constant, so the ground-state variational principle maximizes the mean ferromagnetic bond correlation among compatible many-body states. Crucially, the estimate uses correlations attained by this \emph{zero-temperature thermal state} $\omega_N$, rather than a separately optimized two-particle state. Together with the local second moment this gives the following result.

\DEFTHM{thm:FQbound}For the complete-graph ground-state ensemble specified above, with $L=1$ and $J_1=-J<0$,
\begin{equation}
\mathcal F_Q[\varrho,h_1]\ge\mathcal F_Q^{(N)}[\varrho,h_1]
=4\bigl[\ex{h_1^2}_\varrho-C_N(\varrho)\bigr].\label{eq:FQbound}
\end{equation}
The finite-size estimate obeys
\begin{equation}
\frac{N-1}{N}\mathcal F_Q[\varrho,h_1]\le\mathcal F_Q^{(N)}[\varrho,h_1]\le\mathcal F_Q[\varrho,h_1].\label{eq:rendorelv}
\end{equation}

The upper bound follows from the ground-state variational principle: the energy cannot exceed the separable minimum for its own marginal, while the lower bound follows from the monotonicity of QFI under partial trace and the inequality bounding QFI by four times the variance. The proof, equivalent energy expression, and spectral-range error bound are given in Supplemental Material~\ref{app:thm:FQbound}\footnote{See Supplemental Material for the two-colorable case of Theorem~\ref{thm:equality_holds} (SM1), the derivation of Tables~\ref{tab:max} and \ref{tab:min} (SM2), the proof of Theorem~\ref{thm:FQbound} (SM4), and additional applications. References cited only in the Supplemental Material are Refs.~\cite{Toth2022Uncertainty,Leka2013Some,Petz2014,Toth2015Evaluating,toth2026relationsB,Toth2025QuantumWasserstein,Yu2022Quantum-Inspired,Yu2021AComplete,Gois2023Uncertainty,Moran2024Uncertainty,Xu2024Bounding,Sorensen2001Many-particle,Toth2007Optimal,Vitagliano2011Spin,Vitagliano2014Spin,Vitagliano2025sudsqueezingmany,Caves2002Unknown,Christandl2007One-and-a-Half,Coffman2000Distributed,Osborne2006General,Sorensen2001Entanglement,takahashi1999thermodynamics}}
\nocite{Toth2022Uncertainty,Leka2013Some,Petz2014,Toth2015Evaluating,toth2026relationsB,Toth2025QuantumWasserstein,Yu2022Quantum-Inspired,Yu2021AComplete,Gois2023Uncertainty,Moran2024Uncertainty,Xu2024Bounding,Sorensen2001Many-particle,Toth2007Optimal,Vitagliano2011Spin,Vitagliano2014Spin,Vitagliano2025sudsqueezingmany,Caves2002Unknown,Christandl2007One-and-a-Half,Coffman2000Distributed,Osborne2006General,Sorensen2001Entanglement,takahashi1999thermodynamics}

Equation~\eqref{eq:rendorelv} limits the relative underestimate to $1/N$ when the QFI is nonzero. The absolute error is at most $[\lambda_{\max}(h_1)-\lambda_{\min}(h_1)]^2/N$. Unlike the exact separable-energy formula, this statement concerns the zero-temperature thermal state $\omega_N$ and provides a quantitative approximation to the QFI of its marginal. For $h_1=\sigma_z/2$, the second moment is fixed at $1/4$, so only the correlation is needed. Figure~\ref{fig:FQ_ground}(b) shows the convergence for $N=4,10,60$ and $\varrho=(\openone+|\ex{\sigma_x}_\varrho|\sigma_x)/2$.

The estimated quantity is the QFI of the local state realized by the chosen Hamiltonian and field, not the many-body QFI. The upper bound on the estimation error decreases as $1/N.$ The estimate does not require the zero-temperature thermal state $\omega_N$ itself to be separable. Thus the proposal establishes a controlled correlation-based route to single-particle QFI for the local states accessible through the chosen Hamiltonian, complementing other measurement approaches \cite{Hauke2016Measuring,Apellaniz2017Optimal,MullerRigat2023CertifyingQuantum,Vitale2024Robust,Rath2021Quantum}.

{\it General two-particle reduction.} The ferromagnetic example works because one state realizes the optimal correlation on every bond. For general interactions, the first step is still to optimize a single bond; the additional question is whether its optimum extends to the whole graph. This representability problem is nontrivial even for three qubits: separable states $\varrho_{12}$ and $\varrho_{23}$ need not admit any joint state $\varrho_{123}$ having them as marginals, even if $\operatorname{Tr}_1(\varrho_{12})=\operatorname{Tr}_3(\varrho_{23})$.

We stress that we do not aim to obtain a closed-form expression for the exact separable-energy minimum of every spin Hamiltonian. Such an expression cannot generally be expected for an arbitrary separability optimization problem.

Let us distribute the field term equally among the bonds and define
\begin{equation}
H_{nn'}=\sum_{l=1}^L J_lh_l^{(n)}h_l^{(n')}-\frac{N}{2N_p}\vec B(\vec g^{(n)}+\vec g^{(n')}),\label{eq:Hmn}
\end{equation}
so that $H=\sum_{\langle n,n'\rangle}H_{nn'}$.

\DEFTHM{thm:The minimum for separable states}\textlabel{thm:equality_holds}{\thetheorem}\textlabel{thm:equality_holds_complete_graph}{\thetheorem}For any regular interaction graph,
\begin{equation}
E_{\rm sep}(\varrho)\ge N_p\min_{\varrho_{AB}\in\mathrm{Sep}_2(\varrho)}\ex{H_{AB}}_{\varrho_{AB}}.\label{eq:Hbound}
\end{equation}
Equality holds on two-colorable graphs, whose sites can be divided into two sets so that every bond joins different sets. Within symmetric separable states, the corresponding optimization is exact on every regular graph:
\begin{equation}
E_{\rm symsep}(\varrho)=N_p\min_{\varrho_{AB}\in\mathrm{SymSep}_2(\varrho)}\ex{H_{AB}}_{\varrho_{AB}}.\label{eq:Hbound_sym}
\end{equation}
If $J_l\le0$ for all $l$, then $E_{\rm sep}(\varrho)=E_{\rm symsep}(\varrho)$ on any regular graph.

\begin{figure}
\includegraphics[width=0.65\columnwidth]{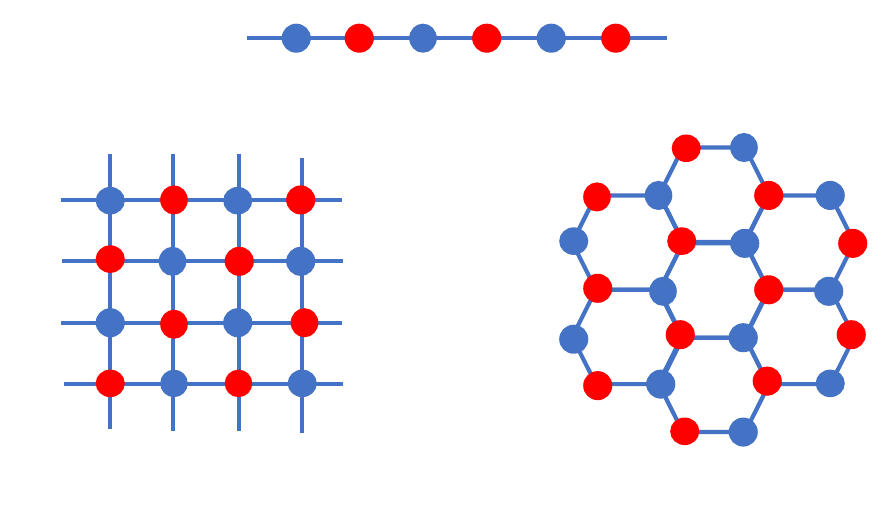}

\vskip-0.2cm
(a)
\\\vskip0.5cm
\includegraphics[width=0.45\columnwidth]{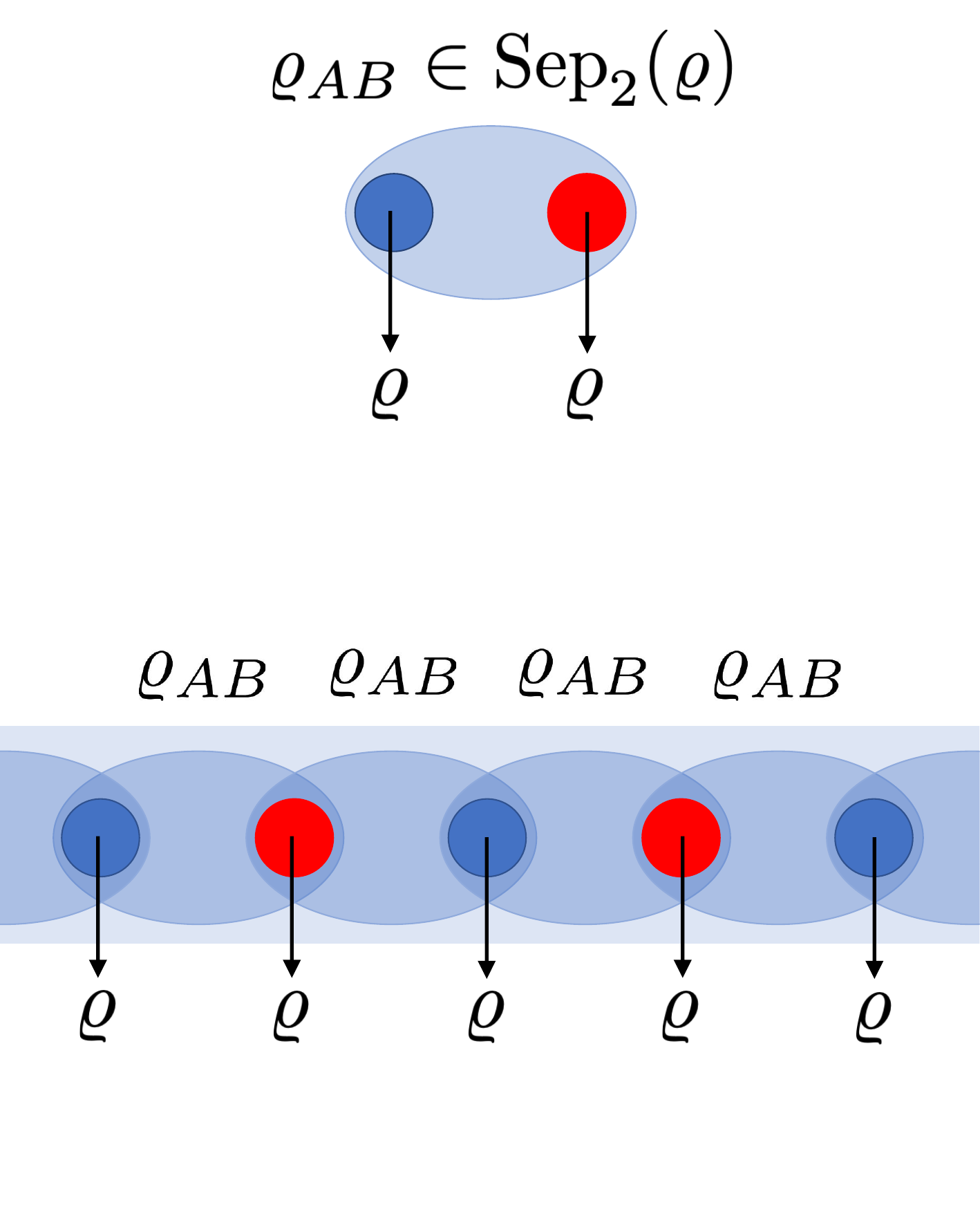}
\includegraphics[width=0.45\columnwidth]{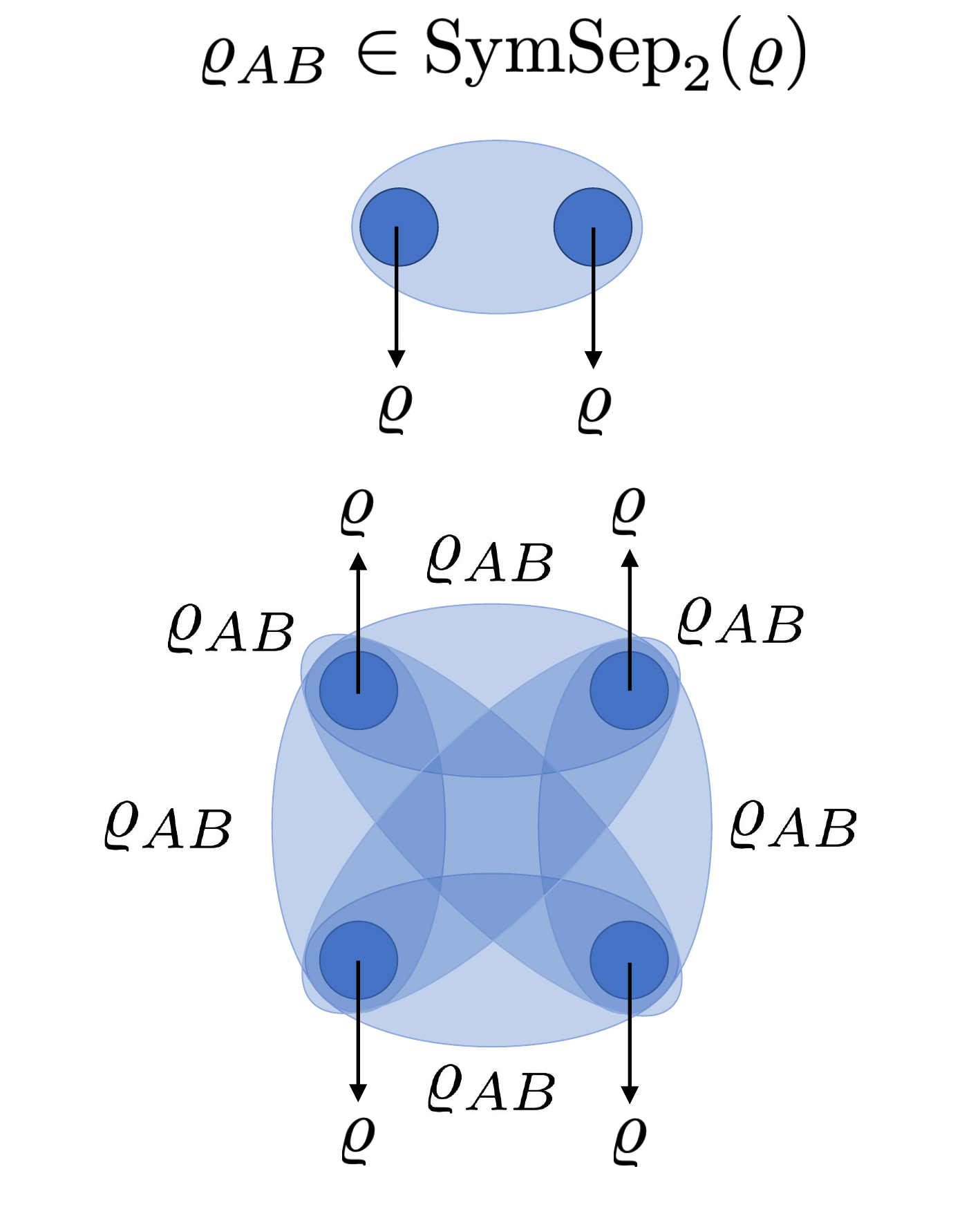}

(b) \hskip3.7cm (c)
\vskip0.2cm

\caption{(a) Spin models on two-colorable graphs. (top) Spin chain. (left) Two-dimensional cubic lattice. (right) Hexagonal lattice.
(b)  $N$-particle separable state with two-particle nearest-neighbor marginals $\varrho_{AB}.$ (top) Two-particle separable state $\varrho_{AB}$ with marginals $\varrho.$ (bottom) A portion of an $N$-particle chain.
(c) $N$-particle symmetric separable state with two-particle nearest-neighbor marginals $\varrho_{AB}$. (top) Two-particle symmetric separable state $\varrho_{AB}$ with single-particle reduced states $\varrho.$ (bottom) Symmetric separable state for $N=4$.
} 
\label{fig:twocolorable}
\end{figure}

{\it Proof.} The lower bound follows by minimizing each bond separately. On a two-colorable graph, each product component $|\Psi_k\rangle\otimes|\Phi_k\rangle$ of an optimal separable two-particle state extends by assigning $|\Psi_k\rangle$ to one sublattice and $|\Phi_k\rangle$ to the other [Fig.~\ref{fig:twocolorable}(a,b)]. Mixing these components reproduces the prescribed marginals and attains the bond minimum throughout the graph; see Supplemental Material~\ref{app:thm:equality_holds}. This includes periodic nearest-neighbor hypercubic lattices with all linear sizes even, for which $N_p=DN$ in $D$ dimensions.

For a symmetric separable optimum, the extension is simply a mixture of $|\Psi_k\rangle\langle\Psi_k|^{\otimes N}$ [Fig.~\ref{fig:twocolorable}(c)] \cite{Korbicz2005Spin}. It works independently of the graph, including the complete graph. If $J_l\le0$, Eq.~\eqref{eq:GMPC_distance_sep2} in Supplemental Material SM2.F shows that the separable bond minimum is attained by a symmetric separable state. The same extension then attains the many-body separable minimum. \qed

A general separable bond state need not have such an extension \cite{Toth2009Spin,Navascues2021Entanglement}; this is why the two saturation conditions differ. Uniform factorized ground states,
with the same pure state at every site, are special cases
of symmetric separable states \cite{Giampaolo2008Theory}.

This gives a second, more general contribution: even when a global extension cannot be constructed, the bond optimization in \THM{thm:The minimum for separable states}\ supplies a systematic many-body energy bound. The distinction between a bound and an attainable minimum is fixed by the graph and the allowed class of states, not by the numerical method used to solve the bond problem.

The computational problem now involves only two particles. For qubits it is a semidefinite program, because positivity under partial transposition is equivalent to separability \cite{Vandenberghe1996Semidefinite,Horodecki1997Separability,Peres1996Separability}. In higher local dimensions, systematic hierarchies give lower bounds \cite{Doherty2002Distinguishing,Doherty2004Complete,Doherty2005Detecting}. Tables~\ref{tab:max} and \ref{tab:min} in End Matter Appendix~\ref{Appendix A: Tables} collect analytical expressions and bounds, with derivations in Supplemental Material~\ref{app:convex_concave}.

{\it Further exact results and multipartite bounds.} The sign of the interaction selects the relevant correlation extremum. Ferromagnetic couplings favor maximal correlations, leading to the QFI expression above; antiferromagnetic couplings favor minimal correlations. For the symmetric-separable models below, a suitable decomposition attains these minima while fixing the relevant single-particle expectation values, so the energy depends on local means rather than QFI. On complete graphs with $J_l>0$ and $L=1$ (Ising-like) or $L=2$ (XY-like),
\begin{equation}
E_{\rm symsep}(\varrho)=N_p\sum_{l=1}^{L}J_l\ex{h_l}_\varrho^2-N\vec B\ex{\vec g}_\varrho,\qquad L=1,2.\label{eq:antif_exact}
\end{equation}
For these two cases, a pure-state decomposition can keep the relevant expectation values fixed at their values in $\varrho$ in every component, attaining the smallest symmetric-separable bond correlation [Table~\ref{tab:min}(b)]. With more positive interaction terms the expression is generally only a lower bound. Details are in End Matter Appendix~\ref{Appendix B: Antiferromagnetic systems} and Supplemental Material~\ref{sec:antiferro}.

The local marginals can also differ. In an even periodic ferromagnetic spin-$1/2$ Heisenberg chain, take alternating marginals $\varrho$ and $\sigma$, $J_1=J_2=J_3=-J<0$, and $h_l=j_l=\sigma_l/2$. The two-qubit correlation maximum in Table~\ref{tab:max}(d) yields
\begin{align}
E_{\rm sep}(\varrho,\sigma)&=-JN\left[\frac12F(\varrho,\sigma)-\frac14\right]\nonumber\\
&\quad-\frac N2\vec B\bigl(\ex{\vec\sigma}_\varrho+\ex{\vec\sigma}_\sigma\bigr).\label{eq:heis_exact}
\end{align}
Here $F(\varrho,\sigma)=\mathrm{Tr}[\sqrt{\sqrt\varrho\,\sigma\sqrt\varrho}]^2$ is the Uhlmann--Jozsa fidelity. It determines how strongly neighboring spins can align while retaining the prescribed marginals. Unlike the Ising-like case, this optimization involves different local states at the two ends of each bond. Fidelity quantifies the corresponding loss of maximal exchange correlation: for identical marginals it equals one, while different local states generally permit less alignment. Supplemental Material~\ref{sec:ferro} gives the derivation and correlation-based fidelity estimates.

Finally, the method can retain entanglement inside contiguous blocks while bounding correlations between them. It gives energy bounds for states separable across a fixed partition into $k$-particle blocks, as described in End Matter Appendix~\ref{Appendix C: Multipartite entanglement} and Supplemental Material~\ref{sec:multi}. These bounds test separability across that partition; they do not by themselves certify unrestricted entanglement depth.

{\it Conclusions.} We show that minimizing the energy over separable many-body states with prescribed single-particle marginals can be reduced to a two-particle optimization. For ferromagnetic Ising-like interactions on any regular graph, this gives an exact expression involving the local QFI. For more general interactions, the two-particle optimization gives the exact separable-energy minimum on two-colorable regular graphs and the exact symmetric-separable minimum on any regular graph. We also show that ground-state correlations of fully connected ferromagnetic systems can be used to estimate local QFI with a finite-size error bounded by $O(1/N)$. Further exact results for antiferromagnetic models and ferromagnetic Heisenberg chains involve local expectation values and quantum fidelity.

We thank A. Ac\'{i}n for insightful discussions, especially for suggesting that we examine multipartite entanglement.  We thank I.~Apellaniz, M.~Eckstein, F. Fr\"owis, M.~Gessner, C.~Klempt, J.~Ko\l ody\'nski, M.~W.~Mitchell, M.~Mosonyi, G.~Muga, J.~Siewert, Sz.~Szalay, K. \.Zyczkowski, T.~V\'ertesi, G. Vitagliano, and D. Virosztek  for discussions. We acknowledge the support of the EU through COST Action CA23115, the MCIN/AEI/10.13039/501100011033 under Grant No.~PCI2022-132947 (co-funded by European Union NextGenerationEU/PRTR), and Grants No.~PID2021-126273NB-I00 (co-funded by 'ERDF A way of making Europe') and No.~PID2025-171880NB-I00.
We thank the Basque Government (Grants No.~IT1470-22 and IT1887-26), and the National Research, Development and Innovation Office NKFIH (Grant No.~2019-2.1.7-ERA-NET-2021-00036, Advanced Grant No. 152794).
We thank the "Frontline" Research Excellence Programme of the NKFIH (Grant No.~KKP133827).~This project has received funding from the European Union's Horizon 2020 Research and Innovation Programme under Grant Agreement No.~731473 and 101017733 (QuantERA MENTA, QuantERA QuSiED).
We thank Project No.~TKP2021-NVA-04, which has been implemented with the support provided by the Ministry of Innovation and Technology of Hungary from the National Research, Development and Innovation Fund, financed under the TKP2021-NVA funding scheme.
We thank the Quantum Information National Laboratory of Hungary. 

\bibliography{Bibliography2}

\vskip0.5cm


\appendix

\setcounter{equation}{0}
\renewcommand{\theequation}{A\arabic{equation}}
\renewcommand{\theHequation}{A\arabic{equation}}

\begin{center}
{\large \bf End Matter}
\end{center}

\section{Appendix A: Tables~\ref{tab:max} and \ref{tab:min}}
\textlabel{Appendix A: Tables}{A}

Tables~\ref{tab:max} and \ref{tab:min} summarize analytical expressions and bounds for two-particle correlation extrema at fixed marginals. Their derivations are given in Supplemental Material~\ref{app:convex_concave} \cite{Note1}.

\paragraph{Analytic solution for qubits.}
For qubits, the optimization problems in Tables I and II
admit explicit solutions for arbitrary $L$. 
Let
\begin{equation}
\varrho=\frac{1}{2}(\openone+\vec r\cdot\vec\sigma),
\qquad
h_l=a_l\openone+\vec b_l\cdot\vec\sigma,
\end{equation}
and define
\begin{equation}
M=\sum_{l=1}^L\vec b_l\vec b_l^{\,T},\qquad
A(\vec r)=\sum_{l=1}^L(a_l+\vec b_l\cdot\vec r)^2.
\end{equation}
The corresponding two-particle correlation extrema are
\begin{equation}
\begin{aligned}
&\max_{\varrho_{AB}\in{\rm Sep}_2(\varrho)}
\sum_{l=1}^L\langle h_l\otimes h_l\rangle_{\varrho_{AB}}\\
&\qquad\qquad=A(\vec r)+\lambda_{\max}(M)(1-|\vec r|^2),\\
&\min_{\varrho_{AB}\in{\rm SymSep}_2(\varrho)}
\sum_{l=1}^L\langle h_l\otimes h_l\rangle_{\varrho_{AB}}\\
&\qquad\qquad=A(\vec r)+\lambda_{\min}(M)(1-|\vec r|^2).
\end{aligned}\label{eq:qubitformulas}
\end{equation}
The maximum of the first equation is also attained within ${\rm SymSep}_2(\varrho)$.
The proof is given in \THM{thm:qubitcase} and \ref{suppl:F. Inequalities with two-body correlations}.

\begin{table}[htbp]
\centering
\begin{tabular}{|c|c|c|c|}
\hline
Line & $L$&Set of states&$\max\sum_{l=1}^L\ex{h_l\otimes h_l}_{\varrho_{AB}}$\\
\hline
(a)\phantom{*}&$1$ & $\mathrm{Sep}_2,\mathrm{SymSep}_2$ & $\ex{h_1^2}_{\varrho}-{\mathcal F}_Q[\varrho,h_1]/4$\\
(b)\phantom{*}&$\ge2$ & $\mathrm{Sep}_2,\mathrm{SymSep}_2$ &$\le\sum_l(\ex{h_l^2}_{\varrho}-{\mathcal F}_Q[\varrho,h_l]/4)$ \\
(c)\phantom{*}&$1$& $\mathcal D_2, d=2$ & $\ex{h_1^2}_{\varrho}-I^{\rm WY}_{\varrho}(h_1)$ \\
(d)$^*$&$3$& $\mathrm{Sep}_2(\varrho,\sigma), d=2$ & $F(\varrho,\sigma)/2-1/4$ \\
(e)\phantom{*}&$\ge1$&  \begin{tabular}{@{}c@{}}$\mathrm{Sep}_2,\mathrm{SymSep}_2,$\\ $d=2$ \end{tabular}& \begin{tabular}{@{}c@{}} Exact formula,\\\EQ{eq:qubitformulas} \end{tabular}\\
\hline
\end{tabular}
\caption{Analytical expressions and bounds for the maxima of two-particle correlations at fixed marginals. We assume that both single-particle reduced states of $\varrho_{AB}$ are  $\varrho.$ Here, $\mathrm{Sep}_N$ and $\mathrm{SymSep}_N$ denote the sets of $N$-particle separable states and symmetric separable states, respectively. $\mathcal D_N$ is the set of $N$-particle quantum states. \; $^*$~indicates that $h_1=j_x, h_2=j_y, h_3=j_z,$ and the reduced states are $\varrho$ and $\sigma.$ For the derivation of the bounds, see \ref{app:convex_concave} \cite{Note1}. 
Line (a) can be used to give a tight lower bound for separable states and symmetric separable states for ferromagnetic Ising and Ising-like models. 
Line (b) can be used to give a not necessarily tight lower bound for separable states and symmetric separable states for ferromagnetic XY-like and Heisenberg-like models. Line (c) can be used to give the energy minimum for two-particle quantum states for the same system, or a lower bound on the energy minimum for a chain with an even number of qubits. Line (d) can be used to give a lower bound for separable states for Heisenberg chains. (e) For qubits, an exact formula is given in \EQ{eq:qubitformulas} for arbitrary $L.$}
\label{tab:max}
\end{table}

\begin{table}[htbp]
\centering
\begin{tabular}{|c|c|c|c|}
\hline
Line &$L$&\cphantom{Set of states}{$\mathrm{Sep}_2,\mathrm{SymSep}_2$}&\cphantom{$\min\sum_{l=1}^L\ex{h_l\otimes h_l}_{\varrho_{AB}}$}{$\le\sum_l\ex{h_l^2}_{\varrho}-{\mathcal F}_Q[\varrho,h_l]/4$}\\
\hline
(a)\phantom{*} &$\ge1$&$\mathrm{Sep}_2$&$\le\sum_l\ex{h_l}_{\varrho}^2$\\
(b)\phantom{*} &$1,2$&$\mathrm{SymSep}_2$&$=\sum_l\ex{h_l}_{\varrho}^2$\\
(c)\phantom{*} &$\ge3$&$\mathrm{SymSep}_2$&$\ge\sum_l\ex{h_l}_{\varrho}^2$\\
(d)\phantom{*}&$\ge1$&  \begin{tabular}{@{}c@{}} $\mathrm{SymSep}_2,$\\ $d=2$\end{tabular} & \begin{tabular}{@{}c@{}}  Exact formula,\\ \EQ{eq:qubitformulas} \end{tabular}\\
\hline
\end{tabular}
\caption{Analytical expressions and bounds for the minima of two-particle correlations at fixed marginals. 
For the notation see the caption of Table~\ref{tab:max}. 
For the derivation of the bounds, see \ref{app:convex_concave} \cite{Note1}. 
Line (b) can be used to give a tight lower bound for symmetric separable states for antiferromagnetic Ising and Ising-like models with $L=1$ and antiferromagnetic $XY$ and $XY$-like models with $L=2.$ Line (c) can be used to give a not necessarily tight lower bound for symmetric separable states for antiferromagnetic Heisenberg-like models. (d) For qubits, an exact formula is given in  \EQ{eq:qubitformulas} for arbitrary $L.$}
\label{tab:min}
\end{table}

\section{Appendix B: Antiferromagnetic systems}
\textlabel{Appendix B: Antiferromagnetic systems}{B}
\renewcommand{\theequation}{B\arabic{equation}}
\renewcommand{\theHequation}{B\arabic{equation}}
\setcounter{equation}{0}

The complete interaction graph has
\begin{equation}
N_p=N(N-1)/2.\label{eq:Np}
\end{equation}

For an antiferromagnetic Ising or Ising-like system ($J_1>0$, $L=1$) on a fully connected lattice with a symmetric ground state, the energy minimum for symmetric separable states is given by a formula involving expectation values of single-particle operators in the reduced state. This remains true even for the XY Hamiltonian and XY-like Hamiltonians with $J_1,J_2>0$ and $L=2$, where $h_l^{(n)}$ are arbitrary single-particle operators acting on spin $n.$ See Table~\ref{tab:min}(b) and \ref{sec:antiferro}.

\section{Appendix C: Multipartite entanglement} 
\textlabel{Appendix C: Multipartite entanglement}{C}
\setcounter{equation}{0}
\renewcommand{\theequation}{C\arabic{equation}}
\renewcommand{\theHequation}{C\arabic{equation}}

For a periodic nearest-neighbor spin chain, fix the partition $\mathcal P_k=(1,\ldots,k)|(k+1,\ldots,2k)|\cdots$ into contiguous $k$-particle blocks \cite{Guhne2005Multipartite}. Let $\mathcal E_N^{(k)}(\varrho)$ be the states separable across this partition with every single-particle marginal equal to $\varrho$, and denote their minimum energy by $E_{\mathcal P_k}(\varrho)$. This is a subset of all $k$-producible states: the bounds below exclude separability across $\mathcal P_k$, but do not by themselves certify entanglement depth greater than $k$.
The Hamiltonian for a single partition block reads
\be
H_{k\text{-part}}=\sum_{n=1}^{k-1} H_{n,n+1},\label{eq:Hhpart}
\ee 
where $H_{nn'}$ is defined in \EQ{eq:Hmn}. For this fixed partition we obtain the following bound.

\begin{samepage}
\DEFTHM{thm:esepbound_kprod}For $N$ divisible by $k$ and $1\leq k<N$, the minimum energy over $\mathcal E_N^{(k)}(\varrho)$ for a periodic ferromagnetic spin chain with $L=1$ and $J_1=-J<0$ satisfies
\begin{align}
&E_{\mathcal P_k}(\varrho)\ge\frac N k \min_{\varrho_{k-\rm part}\in\mathcal D_k(\varrho)}\ex{H_{k-\rm part}}_{\varrho_{k-\rm part}} \nonumber\\
&\quad\quad\quad-\frac N k J\left(\ex{h_1^2}_\varrho-\frac1 4 {\mathcal F}_Q[\varrho,h_1]\right)- \frac {N}k\vec B\ex{\vec g}_\varrho.\label{eq:bound2}
\end{align}
\end{samepage}

The proof is given in the Supplemental Material, see \ref{sec:multi} \cite{Note1}.

Note that assuming a tensor product of $k$-particle blocks yields a different bound
\begin{align}
\frac N k \min_{\varrho_{k-\rm part}\in\mathcal D_k(\varrho)}\ex{H_{k-\rm part}}_{\varrho_{k-\rm part}} -\frac N k J\ex{h_1}_\varrho^2- \frac {N}k\vec B\ex{\vec g}_\varrho.\label{eq:secondm}
\end{align}
This expression is always larger than or equal to the right-hand side of \EQ{eq:bound2}.

\begin{figure}[tbp]
\includegraphics[width=0.47\columnwidth]
{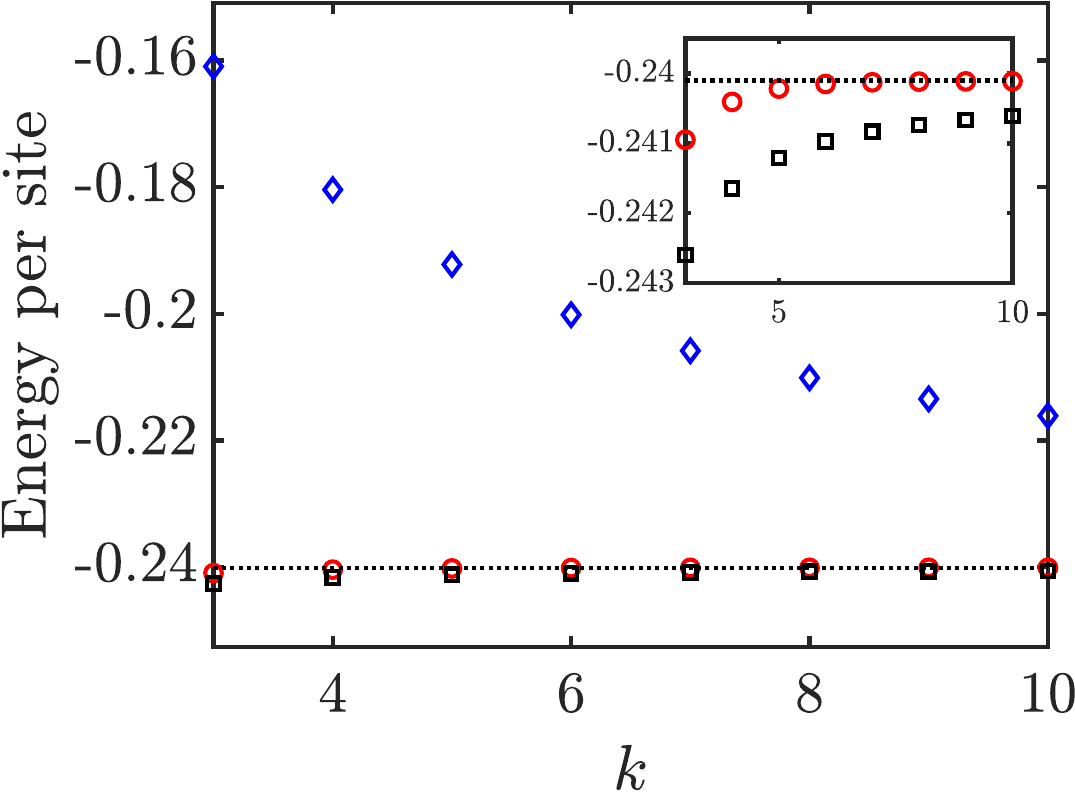}\hspace{0.3cm}
\includegraphics[width=0.47\columnwidth]
{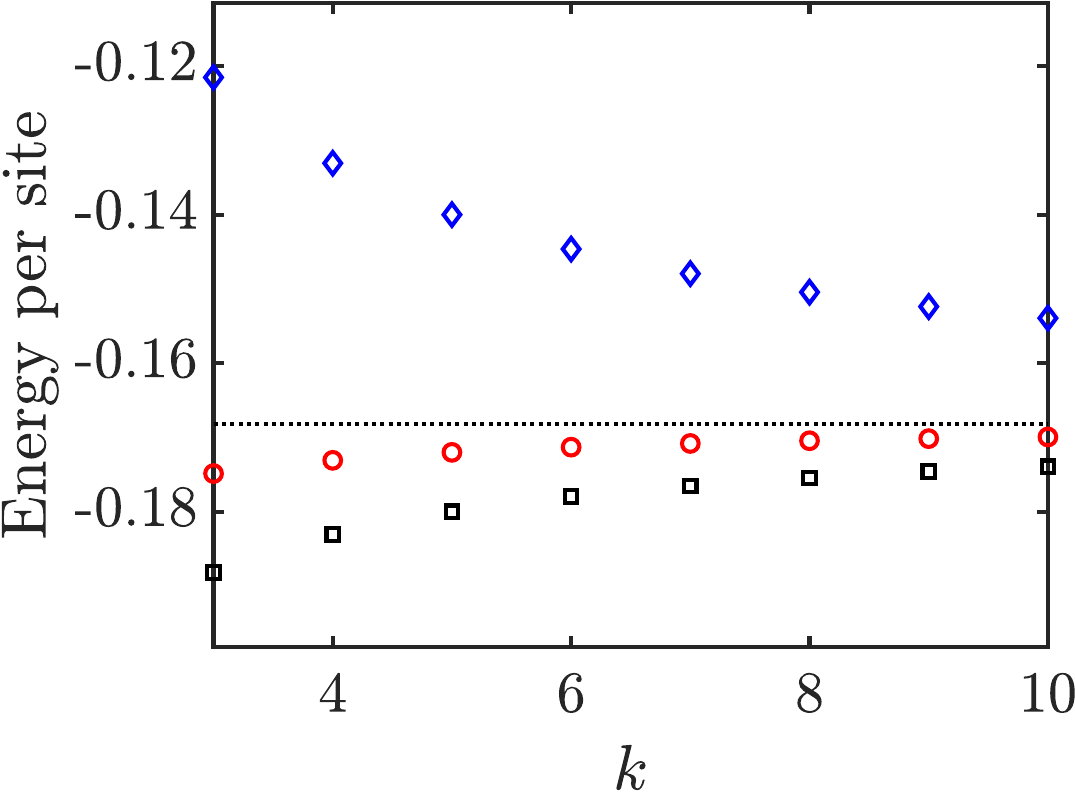}
\hspace*{0.60cm} (a) \hspace{3.8cm} (b)
\caption{Energy bounds for states separable across the fixed partition $\mathcal P_k$ into contiguous $k$-particle blocks for a periodic Ising spin chain with $J=1$ for (a) $\ex{j_x}_{\varrho}=0.1$ and (b) $0.3.$ (diamond)  Bound for a tensor product of $k$-particle states given in \EQ{eq:secondm}.
(circle) Lower bound for states separable across $\mathcal P_k$, using the single-particle quantum Fisher information, \EQ{eq:bound2}.
(square) Lower bound for the energy per particle with the Wigner--Yanase skew information. 
(dotted) Energy minimum per qubit for infinite systems \cite{Pfeuty1970The}. 
} \label{fig:Ising_chain_ground_maintext}
\end{figure}

\clearpage

\renewcommand{\theHfigure}{S\arabic{figure}}
\renewcommand{\theHtable}{S\arabic{table}}
\renewcommand{\theHequation}{S\arabic{equation}}
\renewcommand{\theHtheorem}{S\arabic{theorem}}
\renewcommand{\theHsection}{SM\arabic{section}}

\renewcommand{\thefigure}{S\arabic{figure}}
\renewcommand{\thetable}{S\arabic{table}}
\renewcommand{\theequation}{S\arabic{equation}}
\renewcommand{\thetheorem}{S\arabic{theorem}}
 \renewcommand{\thesection}{SM\arabic{section}}

\stepcounter{myfigure}
\stepcounter{mytable}
\stepcounter{myequation}
\stepcounter{mytheorem}
\stepcounter{mysection}
\setcounter{page}{1}
\thispagestyle{empty}

\onecolumngrid



\begin{center}
{\large \bf Supplemental Material for \\``Quantum Fisher Information and Separable Energy Minima in Spin Systems''}

\bigskip
G\'eza T\'oth$^{1,2,3,4,5}$ and J\'ozsef Pitrik$^{6,7,5}$ 

{$^1$\it\small Theoretical Physics,
University of the Basque Country
UPV/EHU, P.O. Box 644, E-48080 Bilbao, Spain}

{$^2$\it\small EHU Quantum Center, University of the Basque Country UPV/EHU, 
48940 Leioa, 
Spain}

{$^3$\it\small Donostia International Physics Center DIPC,  
20018 San Sebasti\'an, Spain}

{$^4$\it\small IKERBASQUE, Basque Foundation for Science,
E-48009 Bilbao, Spain}

{$^5$\it\small HUN-REN Wigner Research Centre for Physics, H-1525 Budapest, Hungary}

{$^6$\it\small Department of Analysis and Operations Research, Institute of Mathematics, Budapest University of Technology and Economics, 
1111 Budapest, Hungary}

{$^7$\it\small HUN-REN 
R\'enyi Institute of Mathematics, 
1053 Budapest, Hungary}

(Dated: \today)

\medskip
\medskip

\parbox[b][1cm][t]{0.85\textwidth}{\quad
We prove Theorem~\ref{thm:FQbound}.  We give a detailed proof of the two-colorable case of Theorem~\ref{thm:equality_holds}. We derive the bounds in Tables \ref{tab:max} and \ref{tab:min}.   We show how optimal entanglement conditions can be derived from our approach.  We compute the energy minimum for symmetric separable states with a given marginal for antiferromagnetic spin systems. We consider ferromagnetic spin chains and show that the Uhlmann--Jozsa fidelity appears in the formula giving the energy minimum for separable states. With the same techniques, we derive bounds for states separable across fixed partitions into contiguous blocks.}
\bigskip
\bigskip
\bigskip
\bigskip
\bigskip

\end{center}

\bigskip

\twocolumngrid

\section {SM1. Two-colorable case of \THM{thm:equality_holds}}

\textlabel{app:thm:equality_holds}{SM1}

Let us assume that a two-particle separable state $\varrho'_{AB}$ minimizes  the right-hand side of \EQ{eq:Hbound}. Let us now consider the state 
\be
\varrho_{AB}''=F \varrho'_{AB} F,
\ee
where $F$ is the flip operator. Since the two-particle Hamiltonian in \EQ{eq:Hmn} is permutationally invariant, for its expectation value 
\be
\ex{H_{AB}}_{\varrho_{AB}''}=\ex{H_{AB}}_{\varrho_{AB}'}
\ee
holds. Then, we can construct the separable state 
\be
\varrho_{AB}=\frac{1}{2}(\varrho_{AB}'+\varrho_{AB}'')
\ee
that has the same expectation value for $H_{AB},$ and it is permutationally invariant,
\be
\varrho_{AB}=F \varrho_{AB} F.\label{eq:perminv}
\ee
Let us assume that the separable state $\varrho_{AB}$ is decomposed as 
\be
\varrho_{AB}=\sum_l p_l \ketbra{\Psi_l}\otimes\ketbra{\Phi_l}.\label{eq:decomp}
\ee

Let us consider a spin system on a lattice that can be partitioned into two disjoint sublattices in such a way that interacting spins correspond to different sublattices, $\mathcal A$ and $\mathcal B$. \FIG{fig:twocolorable} shows such lattices of some common one- and two-dimensional spin models. Then, the Hamiltonian can be given as
\be
H=\sum_{l=1}^{N_p} H_{a_l,b_l},
\ee
where $H_{a_l,b_l}$ is given in \EQ{eq:Hmn}, $a_l$ and $b_l$ are the spins participating in the $l$th interaction, and $a_l$ and $b_l$ are spins in the $\mathcal A$ and $\mathcal B$ sublattices, respectively. Then, let us consider the separable state
\be
\varrho_{\rm opt}=\sum_l p_l \bigotimes_{n=1}^N \ketbra{\chi_{l,n}}\label{eq:rhoprod}
\ee
with the single-particle states
\be
\ket{\chi_{l,n}}=\begin{cases}
\ket{\Psi_l},&\text{ if }n\text{ is in sublattice }\mathcal A,\\
\ket{\Phi_l}&\text{ if }n\text{ is in sublattice }\mathcal B.
\end{cases}\label{eq:state}
\ee
For the two-particle reduced states for all $n$ and $n'$ connected by an interaction
\be
\varrho_{n,n'}=\varrho_{AB}\label{eq:nnprime}
\ee
holds, where $n$ is in sublattice $\mathcal A$ and $n'$ is in sublattice $\mathcal B$. Thus, the energy is the sum of the expectation values of the two-particle Hamiltonians in $\varrho_{AB}$ as
\begin{equation}
\ex{H}_{\varrho_{\rm opt}}=\sum_{k=1}^{N_p} \ex{H_{a_k,b_k}}_{\varrho_{AB}}=N_p\ex{H_{AB}}_{\varrho_{AB}},
\end{equation}
and the state \EQ{eq:rhoprod} saturates the inequality in \EQ{eq:Hbound}. 
Here $\ex{H_{AB}}_{\varrho_{AB}}$ is the expectation value of the Hamiltonian given in  \EQ{eq:Hmn} for $\varrho_{AB}.$ \qed

For spin chains, \EQS{eq:perminv} and \eqref{eq:nnprime} imply that
\be
\varrho_{n,n+1}=\varrho_{AB},\label{eq:nnprime2}
\ee
where spin $(N+1)$ equals spin $(1)$ due to the periodic boundary condition.

In particular, \THM{thm:equality_holds} applies to a nearest-neighbor hypercubic lattice with periodic boundary conditions and an even number of sites in every direction.

\section{SM2. Convex roof and concave roof of a sum of variances}
\textlabel{app:convex_concave}{SM2}
 
In this section, we summarize important relations concerning convex and concave roofs.  
We also prove all the statements of Tables~\ref{tab:max} and \ref{tab:min}.

\subsection{A. Definitions of \texorpdfstring{$\mathcal I$ and $\mathcal R$}{I and R}}

For reference, if $\varrho=\sum_k\lambda_k|k\rangle\langle k|$, the quantum Fisher information is
\begin{equation}
\mathcal F_Q[\varrho,h]=2\sum_{k,l}\frac{(\lambda_k-\lambda_l)^2}{\lambda_k+\lambda_l}|\langle k|h|l\rangle|^2,\label{eq:FQ}
\end{equation}
where terms with zero denominator are omitted. Its convex-roof representation is used in the main text and below.

We define the convex roof of the sum of variances as \cite{Toth2022Uncertainty}
\begin{equation}
\mathcal I(\varrho,\{h_l\}_{l=1}^{L})= \min_{\{p_k,\ket{\Psi_k}\}}\sum_k p_k \sum_{l=1}^{L} \va{h_l}_{\Psi_k},\label{eq:infsum}
\end{equation}
and its concave roof as
\begin{equation}
\mathcal R(\varrho,\{h_l\}_{l=1}^{L})= \max_{\{p_k,\ket{\Psi_k}\}}\sum_k p_k \sum_{l=1}^{L} \va{h_l}_{\Psi_k}.\label{eq:infsum2}
\end{equation}
The optimization is carried out over all decompositions of $\varrho$ of the form
\begin{equation}
\varrho=\sum_k p_k \ketbra{\Psi_k}, \label{eq:purestatedecomp}
\end{equation}
where $p_k\ge0$ and $\sum_k p_k=1.$ 

\subsection{B. Inequalities for \texorpdfstring{$\mathcal I$ and $\mathcal R$}{I and R} with the variance and the quantum Fisher information} 

Now, we need to know that quantum Fisher information is convex in the state, while the variance is concave. For pure states, we have 
\be
{\mathcal F}_Q[\varrho,h]=4\va{h}_\varrho.
\ee
Based on these, the inequalities \cite{Toth2022Uncertainty}
\begin{subequations}
\begin{align}
&\mathcal I(\varrho,\{h_l\}_{l=1}^{L})\ge\frac1 4\sum_{l=1}^L {\mathcal F}_Q[\varrho,h_l],\label{eq:Iineq}\\
&\mathcal R(\varrho,\{h_l\}_{l=1}^{L})\le\sum_{l=1}^L \va{h_l}_\varrho\label{eq:Rineq}
\end{align}
\end{subequations}
hold. 

In \EQ{eq:Iineq} there is an equality for $L=1$ due to the fact that
the quantum Fisher information is the convex roof of the variance times four  \cite{Toth2013Extremal,Yu2013Quantum,Toth2014Quantum}
\begin{equation}
{\mathcal F}_Q[\varrho,h]=4\min_{\{p_k,\ket{\Psi_k}\}}\;\sum_k p_k \va{h}_{\Psi_k},
\end{equation}
where the optimization is carried out over pure-state decompositions given in \EQ{eq:purestatedecomp}.
Probabilities satisfy $p_k\ge0$ and $\sum_k p_k=1$, and the pure states $\ket{\Psi_k}$ are not assumed to be orthogonal to each other.

In \EQ{eq:Rineq}, there is also an equality for $L=1$ 
due to the fact that the variance is the concave roof of itself  \cite{Toth2013Extremal,Yu2013Quantum,Toth2014Quantum}
\begin{equation}
\va{h}=\max_{\{p_k,\ket{\Psi_k}\}}\;\sum_k p_k \va{h}_{\Psi_k},\label{eq:deffqroof}
\end{equation}
where the optimization is carried out over pure-state decompositions given in \EQ{eq:purestatedecomp}.
In \EQ{eq:Rineq}, there is even an equality for $L=2$, since for any $h_1$ and $h_2$ \cite{Leka2013Some,Petz2014}
\begin{equation}
\va{h_1}+\va{h_2}=\max_{\{p_k,\ket{\Psi_k}\}}\;\sum_k p_k [\va{h_1}_{\Psi_k}+\va{h_2}_{\Psi_k}]
\end{equation}
holds.

\subsection{C. Definitions of \texorpdfstring{$\mathcal I'$ and $\mathcal R'$}{I' and R'}} 

We will now define some quantities alternative to $\mathcal I$ and $\mathcal R.$ They will be given as convex and concave roofs for sums of squares of expectation values.

In order to do that, let us rewrite the definition in \EQ{eq:infsum} as
\begin{align}
\mathcal I(\varrho,\{h_l\}_{l=1}^{L})&= \min_{\{p_k,\ket{\Psi_k}\}}\sum_k p_k \sum_{l=1}^{L}\left[\ex{h_l^2}_{\Psi_k}-\ex{h_l}^2_{\Psi_k}\right]\nonumber\\
&=\bigg\langle\sum_{l=1}^L h_l^2\bigg\rangle_{\varrho}-\max_{\{p_k,\ket{\Psi_k}\}}\sum_k p_k\sum_{l=1}^{L}\ex{h_l}^2_{\Psi_k},
\end{align}
where the optimization is carried out over pure-state decompositions given in \EQ{eq:purestatedecomp}.
Here, we could remove the term with the second moments from the optimization over the decompositions. Similarly, \EQ{eq:infsum2} can be rewritten as 
\begin{align}
\mathcal R(\varrho,\{h_l\}_{l=1}^{L})&= \max_{\{p_k,\ket{\Psi_k}\}}\sum_k p_k \sum_{l=1}^{L}\left[\ex{h_l^2}_{\Psi_k}-\ex{h_l}^2_{\Psi_k}\right]\nonumber\\
&=\bigg\langle\sum_{l=1}^L h_l^2\bigg\rangle_{\varrho}-\min_{\{p_k,\ket{\Psi_k}\}}\sum_k p_k\sum_{l=1}^{L}\ex{h_l}^2_{\Psi_k},
\end{align}
where, again, the optimization is carried out over pure-state decompositions given in \EQ{eq:purestatedecomp}.
Then, we can define a quantity based on $\mathcal I(\varrho,\{h_l\}_{l=1}^{L})$ as
\begin{align}
\begin{split}
\mathcal I'(\varrho,\{h_l\}_{l=1}^{L})&=\bigg\langle\sum_{l=1}^L h_l^2\bigg\rangle_{\varrho}-\mathcal I(\varrho,\{h_l\}_{l=1}^{L})\\
&= \max_{\{p_k,\ket{\Psi_k}\}}\sum_k p_k \sum_{l=1}^{L} \ex{h_l}^2_{\Psi_k},\label{eq:Iprime}
\end{split}
\end{align}
and we can also define a similar quantity based on $\mathcal R(\varrho,\{h_l\}_{l=1}^{L})$ as
\begin{align}
\begin{split}
\mathcal R'(\varrho,\{h_l\}_{l=1}^{L})&=\bigg\langle\sum_{l=1}^L h_l^2\bigg\rangle_{\varrho}-\mathcal R(\varrho,\{h_l\}_{l=1}^{L})\\
&= \min_{\{p_k,\ket{\Psi_k}\}}\sum_k p_k \sum_{l=1}^{L} \ex{h_l}^2_{\Psi_k}.\label{eq:Rprime}
\end{split}
\end{align}

\subsection{D. Expressions with the sum of the squares of expectation values}

Based on \EQ{eq:Iineq} and \EQ{eq:Rineq}, we obtain the following useful relations for $\mathcal I'$ and $\mathcal R'$ 
\begin{subequations}
\begin{align}
&\mathcal I'(\varrho,\{h_l\}_{l=1}^{L})\le\bigg\langle\sum_{l=1}^L h_l^2\bigg\rangle_{\varrho}-\frac1 4\sum_{l=1}^L {\mathcal F}_Q[\varrho,h_l],\label{eq:Ipineq}\\
&\mathcal R'(\varrho,\{h_l\}_{l=1}^{L})\ge\sum_{l=1}^L \ex{h_l}^2_{\varrho},\label{eq:Rpineq}
\end{align}
\end{subequations}
where, based on our previous discussion, \EQ{eq:Ipineq} is saturated for $L=1$ and \EQ{eq:Rpineq} is saturated for $L=1,2,$ that is,
\begin{subequations}\begin{align}
&\mathcal I'(\varrho,\{h_1\})=\bigg\langle h_1^2\bigg\rangle_{\varrho}-\frac1 4 {\mathcal F}_Q[\varrho,h_1],\label{eq:Ipineq2}\\
&\mathcal R'(\varrho,\{h_1\})=\ex{h_1}^2_{\varrho},\label{eq:Rpineq2}\\
&\mathcal R'(\varrho,\{h_l\}_{l=1}^{2})=\sum_{l=1}^2 \ex{h_l}^2_\varrho.\label{eq:Rpineq3}
\end{align}\end{subequations}

For qubits, $\mathcal I'(\varrho,\{h_l\}_{l=1}^{L})$ and $\mathcal R'(\varrho,\{h_l\}_{l=1}^{L})$ can be obtained from explicit formulas. 
Let
\begin{equation}
\varrho=\frac{1}{2}(\openone+\vec r\cdot\vec\sigma),
\qquad
h_l=a_l\openone+\vec b_l\cdot\vec\sigma,
\end{equation}
and define
\begin{equation}
M=\sum_{l=1}^L\vec b_l\vec b_l^{\,T},\qquad
A(\vec r)=\sum_{l=1}^L(a_l+\vec b_l\cdot\vec r)^2.
\end{equation}
\DEFTHM{thm:qubitcase} The quantities defined in \EQ{eq:Iprime} and \EQ{eq:Rprime} are
\begin{align}
\begin{split}
\mathcal I'(\varrho,\{h_l\})
&=A(\vec r)+\lambda_{\max}(M)(1-|\vec r|^2),\\
\mathcal R'(\varrho,\{h_l\})
&=A(\vec r)+\lambda_{\min}(M)(1-|\vec r|^2).
\end{split}
\end{align}

\textit{Proof.} For a pure-state decomposition of $\varrho$, let $\vec n_\alpha$ denote the Bloch vectors and $p_\alpha$ their probabilities, such that $\sum_\alpha p_\alpha\vec n_\alpha=\vec r$ and $|\vec n_\alpha|=1$. The covariance matrix
\begin{equation}
K=\sum_\alpha p_\alpha\vec n_\alpha\vec n_\alpha^{\,T}-\vec r\vec r^{\,T}
\end{equation}
satisfies $K\geq0$ and
$\operatorname{Tr}K=1-|\vec r|^2$.
The objective function becomes
\begin{equation}
A(\vec r)+\operatorname{Tr}(MK).
\end{equation}
Its maximum and minimum are therefore bounded by the largest and smallest eigenvalues of $M$, respectively. Both bounds are attained by a two-state decomposition along an eigenvector corresponding to the respective extremal eigenvalue. This proves the formulas. $\blacksquare$

\subsection{E. Mixed-state decompositions}

We now derive some relevant relations for $\mathcal I'$ and $\mathcal R'$ with an optimization over mixed-state decompositions, rather than considering pure-state decompositions. 

It can be shown that the following relation holds
\begin{align}
\mathcal I'(\varrho,\{h_l\}_{l=1}^{L})&=\max_{\{p_k,\varrho_k\}}\sum_k p_k \sum_{l=1}^{L} \ex{h_l}^2_{\varrho_k}.
\label{eq:Iprimeineq}
\end{align}
In \EQ{eq:Iprimeineq}, there is an optimization over mixed-state decompositions of $\varrho$ of the type
\be
\varrho=\sum_k p_k\varrho_k.\label{eq:decompmixed}
\ee
The relation in \EQ{eq:Iprimeineq} is due to the fact that based on simple convexity arguments, we cannot get a larger value if  we optimize over mixed-state decompositions. In particular, if a convex function is maximized over a convex set, then it attains its maximum at the extreme points of the set. In our case, the convex set is the set of quantum states and the extreme points are pure states \cite{Toth2022Uncertainty}.

We can also consider a minimization over mixed states rather than over pure states, for which we obtain
\begin{align}
\mathcal R'(\varrho,\{h_l\}_{l=1}^{L})&\ge\min_{\{p_k,\varrho_k\}}\sum_k p_k \sum_{l=1}^{L} \ex{h_l}^2_{\varrho_k},\label{eq:Rprimeineq}
\end{align}
where for the decompositions \EQ{eq:decompmixed} holds.
In \EQ{eq:Rprimeineq}, the inequality is due to the fact that on the right-hand side we optimize over a strictly larger set than in the expression on the left-hand side of the inequality \cite{Toth2022Uncertainty}. 

For $L=1$ there is an equality in \EQ{eq:Rprimeineq}. This can be seen since due to the concavity of the variance for any operator $h$ we have \cite{Toth2022Uncertainty}
\be
\sum_k p_k \va{h}_{\varrho_k} \le \va{h}_{\varrho}, 
\ee
which together with \EQ{eq:deffqroof} leads to 
\begin{equation}
\va{h}=\max_{\{p_k,\varrho_k\}}\;\sum_k p_k \va{h}_{\varrho_k},\label{eq:deffqroof_mixed}
\end{equation}
where for the decompositions \EQ{eq:decompmixed} holds.

For $L=2,$ the two sides of the inequality in \EQ{eq:Rprimeineq} are equal to each other, which follows from \EQ{eq:Rpineq3}. In particular, we can have the example of a qubit system with $\varrho=\openone/2$ and $\{h_l\}=\{\sigma_x,\sigma_y\},$ for which the left-hand side of \EQ{eq:Rprimeineq} is $0.$  This is connected to the relations
\begin{align}
\begin{split}
\mathcal I'(\varrho,\{\sigma_x,\sigma_y\})&=1-\ex{\sigma_z}^2_{\varrho},\\
\mathcal R'(\varrho,\{\sigma_x,\sigma_y\})&={\mathcal F}_Q[\varrho,\sigma_z]/4,
\end{split}
\end{align}
since for all pure states we have $\ex{\sigma_x}^2+\ex{\sigma_y}^2=1-\ex{\sigma_z}^2.$ 
The right-hand side of \EQ{eq:Rprimeineq} is zero.
The optimal mixed-state decomposition minimizing the right-hand side of \EQ{eq:Rprimeineq} is $p_1=1, \varrho_1=\varrho.$ 

For $L=3,$  the two sides of the inequality in \EQ{eq:Rprimeineq} are not necessarily equal to each other. We can have the example of a qubit system with $\varrho=\openone/2$ and $\{h_l\}=\{\sigma_x,\sigma_y,\sigma_z\},$
for which the left-hand side of \EQ{eq:Rprimeineq} is $1.$ 
This is connected to the relations valid for all qubit states
\be
\mathcal I'(\varrho,\{\sigma_x,\sigma_y,\sigma_z\})=\mathcal R'(\varrho,\{\sigma_x,\sigma_y,\sigma_z\})=1,
\ee
since for all pure states we have $\ex{\sigma_x}^2+\ex{\sigma_y}^2+\ex{\sigma_z}^2=1.$ 
The right-hand side of \EQ{eq:Rprimeineq} is zero. The optimal mixed-state decomposition minimizing the right-hand side of \EQ{eq:Rprimeineq} is $p_1=1, \varrho_1=\varrho.$ 

\subsection{F. Inequalities with two-body correlations}
\label{suppl:F. Inequalities with two-body correlations}

We summarize the relations of these quantities to two-body correlations. For $\mathcal I'$ we know that \cite{Toth2015Evaluating,Toth2023QuantumWasserstein}
\begin{align}
\mathcal I'(\varrho,\{h_l\}_{l=1}^{L})&= \max_{\varrho_{AB}\in \mathrm{SymSep}_2(\varrho)}\bigg\langle\sum_{l=1}^L h_l\otimes h_l\bigg\rangle_{\varrho_{AB}}\nonumber\\
&= \max_{\varrho_{AB}\in \mathrm{Sep}_2(\varrho)}\bigg\langle\sum_{l=1}^L h_l\otimes h_l\bigg\rangle_{\varrho_{AB}},\label{eq:Ibar}
\end{align}
where $\mathrm{SymSep}_2(\varrho)$ denotes symmetric separable states of two qudits with marginals $\varrho.$ Table~\ref{tab:max}(a) is based on \EQS{eq:Ipineq2} and \eqref{eq:Ibar}, Table~\ref{tab:max}(b) is based on \EQS{eq:Ipineq} and \eqref{eq:Ibar}. The general separable set contains the symmetric separable set, so its maximum could in principle be larger. Equation~\eqref{eq:Ibar} shows that the two maxima coincide. Table~\ref{tab:max}(c) is based on
\be
\max_{\varrho_{AB}\in\mathcal D_2(\varrho)}\ex{h\otimes h}_{\varrho_{AB}}=\ex{h^2}_{\varrho}-I^{\rm WY}_{\varrho}(h),
\ee
proved for two qubits in \REF{toth2026relationsB}. Table~\ref{tab:max}(e) is based on \THM{thm:qubitcase}  and \EQ{eq:Ibar}.

Next, we will prove the second equality in \EQ{eq:Ibar}. Let us assume that
\begin{equation}\label{eq:sep_states}
\varrho_{AB}=\sum_{k}p_{k}\ketbra{\Psi_k}\otimes\ketbra{\Phi_k}.
\end{equation} 
We can rewrite the optimization problem in \EQ{eq:Ibar} for separable states as
\begin{align}
&\quad\quad\quad\quad\max \quad\left\langle\sum_{l=1}^L h_l \otimes h_l\right\rangle_{\varrho_{AB}}\nonumber\\
&\quad\textrm{s.~t. }
\varrho_{AB}\in \mathrm{Sep}_2,\nonumber\\
& \quad\quad\quad{\rm Tr}_2(\varrho_{AB})=\varrho,\nonumber\\
& \quad\quad\quad{\rm Tr}_1(\varrho_{AB})=\varrho,\nonumber\\
&\quad=
\frac1 2  \max_{\{p_k,\ket{\Psi_k},\ket{\Phi_k}\}} \sum_{l=1}^L  \sum_k p_k \bigg[\ex{h_l}^2_{\Psi_k} +  \ex{h_l}^2_{\Phi_k}\nonumber\\
&\quad\quad\quad\quad-\left(\ex{h_l}_{\Psi_k}-\ex{h_l}_{\Phi_k}\right)^2 \bigg]\nonumber\\
&\quad= \max_{\{p_k,\ket{\Psi_k}\}} \sum_{l=1}^L \sum_k p_k \ex{h_l}^2_{\Psi_k},
\label{eq:GMPC_distance_sep2}
\end{align}
where for the second optimization the decomposition of $\varrho_{AB}$ is given as \EQ{eq:sep_states}.
For the third optimization, the condition with $\varrho$ is given in \EQ{eq:purestatedecomp}. Note that the expression on the left-hand side of \EQ{eq:GMPC_distance_sep2} is maximized by $\ket{\Psi_k}=\ket{\Phi_k}$ for all $k,$ and in this case $\left(\ex{h_l}_{\Psi_k}-\ex{h_l}_{\Phi_k}\right)^2=0.$ 
Thus, the expression is maximized by a symmetric separable state, which ends our proof. \qed

An analogous statement holds if we minimize $\sum_l (h_l\otimes\openone-\openone\otimes h_l)^2$ over bipartite separable states with given marginal $\varrho,$ where the minimum is the same for symmetric separable states \cite{Toth2023QuantumWasserstein}.

We also know that \cite{Toth2015Evaluating,Toth2023QuantumWasserstein}
\begin{align}
\mathcal R'(\varrho,\{h_l\}_{l=1}^{L})&= \min_{\varrho_{AB}\in \mathrm{SymSep}_2(\varrho)}\bigg\langle\sum_{l=1}^L h_l\otimes h_l\bigg\rangle_{\varrho_{AB}}\nonumber\\
&\ge \min_{\varrho_{AB}\in \mathrm{Sep}_2(\varrho)}\bigg\langle\sum_{l=1}^L h_l\otimes h_l\bigg\rangle_{\varrho_{AB}}.\label{eq:Rpbound}
\end{align}
Here the minimum for separable states can be smaller than the minimum for symmetric separable states.

The case of the optimization over symmetric separable states is considered in Table~\ref{tab:min}(b-c).
Table~\ref{tab:min}(b) is based on \EQS{eq:Rpineq2}, \eqref{eq:Rpineq3}, and \eqref{eq:Rpbound}.
Table~\ref{tab:min}(c) is based on \EQS{eq:Rpineq} and \eqref{eq:Rpbound}.  Table~\ref{tab:min}(d) is based on \THM{thm:qubitcase}  and \EQ{eq:Rpbound}.

For the minimization over general (not necessarily symmetric) separable states
\be
\min_{\varrho_{AB}\in \mathrm{Sep}_2(\varrho)}\bigg\langle\sum_{l=1}^L h_l\otimes h_l\bigg\rangle_{\varrho_{AB}}\le\sum_{l=1}^L\ex{h_l}^2_{\varrho}\label{eq:sepbound1}
\ee
holds. The minimum over separable states cannot exceed the expectation value for $\varrho_{AB}=\varrho\otimes\varrho$, which is a feasible separable state. Table~\ref{tab:min}(a) is based on \EQ{eq:sepbound1}.

For two qubits we have \cite{Toth2025QuantumWasserstein}
\be
\max_{\varrho_{AB}\in\mathrm{Sep}_2(\varrho,\sigma)} \sum_{l=x,y,z}\langle j_l \otimes j_l\rangle_{\varrho_{AB}}=\frac 1 2 F(\varrho,\sigma)-\frac 1 4,\label{eq:fid111}
\ee
where $F(\varrho,\sigma)$ is the quantum fidelity. $\mathrm{Sep}_2(\varrho,\sigma)$ denotes the set of bipartite separable states with marginals $\varrho$ and $\sigma.$ Table~\ref{tab:max}(d) is based on \EQ{eq:fid111}.

\subsection{G. Other examples of an optimization over two copies}

We note that similar ideas can be used for the $N$-representability problem of pure states and rank-constrained optimization \cite{Yu2022Quantum-Inspired,Yu2021AComplete}. Moreover, there are optimization problems in which the marginal is not constrained \cite{Gois2023Uncertainty,Moran2024Uncertainty,Xu2024Bounding}.  In this case, we obtain
\be
\max_{\varrho}\mathcal R'(\varrho,\{h_l\}_{l=1}^{L}) = \max_{\varrho}\mathcal I'(\varrho,\{h_l\}_{l=1}^{L}) =\max_{\varrho}\sum_{l=1}^{L}\langle h_l\rangle_\varrho^2.  \ee
The maximum is attained in all three cases on pure states. For the minima, we obtain
\begin{align}
\begin{split}
&\min_{\varrho}\mathcal R'(\varrho,\{h_l\}) = \min_{\varrho}\mathcal I'(\varrho,\{h_l\}) \\
&\qquad= \min_{|\Psi\rangle} \sum_{l=1}^L\langle h_l\rangle_\Psi^2\ge\min_{\varrho} \sum_{l=1}^L\langle h_l\rangle_{\varrho}^2. 
\end{split}
\end{align}
When the optimization over $\varrho$ is carried out in the same
direction as the optimization defining $\mathcal I'$ or $\mathcal R'$,
the constraint on the marginal can simply be omitted from the
corresponding optimization problem. Thus, $\max_{\varrho}\mathcal I'$
and $\min_{\varrho}\mathcal R'$ can be obtained by optimizing directly
over all admissible bipartite states, without fixing their marginal.

\section{SM3. Entanglement conditions}
\textlabel{sec:entcond}{SM3}

The energy comparison in Fig.~\ref{fig:FQ_ground}(a) uses the unrestricted two-particle lower bound
\begin{equation}
\min_{\varrho_N\in\mathcal D_N(\varrho)}\ex{H}_{\varrho_N}\ge N_p\min_{\varrho_{AB}\in\mathcal D_2(\varrho)}\ex{H_{AB}}_{\varrho_{AB}}=:E_{\rm LB}(\varrho),\label{eq:Hbound1}
\end{equation}
where $\mathcal D_N(\varrho)$ contains all $N$-particle states with the prescribed single-particle marginals. Each bond marginal is a feasible two-particle state, which proves the bound. For the ferromagnetic Ising-like interaction ($L=1$),
\begin{equation}
E_{\rm LB}(\varrho)=-JN_pC_{\max}(\varrho)-N\vec B\ex{\vec g}_\varrho,\label{eq:EL}
\end{equation}
with $C_{\max}(\varrho)=\max_{\varrho_{AB}\in\mathcal D_2(\varrho)}\ex{h_1\otimes h_1}_{\varrho_{AB}}$.
For qubits, Table~\ref{tab:max}(c) gives $C_{\max}(\varrho)=\ex{h_1^2}_\varrho-I^{\rm WY}_\varrho(h_1)$, where the Wigner--Yanase skew information is
\begin{equation}
I^{\rm WY}_\varrho(h)=\mathrm{Tr}(h^2\varrho)-\mathrm{Tr}(h\sqrt\varrho h\sqrt\varrho).\label{eq:WY}
\end{equation}
If an unconstrained ground state has marginal $\varrho$ at every site, this lower bound and the separable threshold bracket its energy as in Eq.~\eqref{eq:EsepEEL}. Only the separable threshold supplies an entanglement criterion.

We present entanglement conditions for bipartite quantum states. Many of the conditions are optimal since they detect all entangled states that can be detected based on single-particle density matrices and the correlations given.

\subsection{A. Two-particle systems}

Next, we formulate an entanglement condition for the two-particle density matrix $\varrho_{AB}.$

\DEFTHM{thm:optimal}
Let us consider the set of quantum states with non-negative correlations
\be
\ex{h\otimes h}_{\varrho_{AB}}-\ex{h}_{\varrho}^2\ge0,\label{eq:corr}
\ee
where $h$ is a single-particle Hamiltonian.
For separable states satisfying \EQ{eq:corr},
\be
\ex{h\otimes h}_{\varrho_{AB}}\le\ex{h^2}_{\varrho}-\frac1 4 {\mathcal F}_Q[\varrho,h]\label{eq:entcond}
\ee
holds. Any state fulfilling \EQ{eq:corr} and violating the criterion given in \EQ{eq:entcond} is entangled.
The criterion given in \EQ{eq:entcond} is optimal, since it detects all entangled $\varrho_{AB}$ within this set that can be detected based on knowing $\ex{h\otimes h}_{\varrho_{AB}}$ and $\varrho.$

{\it Proof.}   Due to \EQ{eq:corr}, for the set of states we consider, the minimum of the left-hand side of \EQ{eq:entcond} is $\ex{h}_{\varrho}^2,$ and it is taken, for instance, for the product state $\varrho\otimes\varrho.$ 
Due to Table~\ref{tab:max}(a), the inequality in \EQ{eq:entcond} holds for separable states, and there are some separable states that saturate the inequality. $\qed$

Note that \THM{thm:optimal} is true for any $h$ of any dimension, including $d>2.$ This is a surprise, as optimal 
entanglement conditions are difficult to formulate when the dimension of the subsystems is larger than two. 

\DEFTHM{thm:optimal2}For symmetric states, the entanglement condition 
\be
\ex{h^2}_{\varrho}-\va{h}_{\varrho}\le \ex{h\otimes h}_{\varrho_{AB}}\le\ex{h^2}_{\varrho}-\frac1 4 {\mathcal F}_Q[\varrho,h]\label{eq:entcond2}
\ee
is optimal, since it detects all symmetric entangled states $\varrho_{AB}$ that can be detected based on knowing $\ex{h\otimes h}_{\varrho_{AB}}$ and $\varrho.$ 

{\it Proof.}  Due to Table~\ref{tab:min}(b), the first inequality in \EQ{eq:entcond2} holds and it can also be saturated by a symmetric separable state. Due to Table~\ref{tab:max}(a), the second inequality in \EQ{eq:entcond2} holds and it can also be saturated by a symmetric separable state. $\qed$

We present a further entanglement condition with two correlations.

\DEFTHM{thm:optimal3}For symmetric separable states
\be
\sum_{l=1}^2\ex{h_l}_{\varrho}^2\le \sum_{l=1}^2\ex{h_l\otimes h_l}_{\varrho_{AB}}\label{eq:entcond3}
\ee
holds. In \EQ{eq:entcond3}, the bound for symmetric separable states is tight. Any symmetric state that violates the criterion in \EQ{eq:entcond3} is entangled.

{\it Proof.}  See Table~\ref{tab:min}(b). $\qed$

\DEFTHM{thm:optimalfid}Let us consider the set of two-qubit quantum  states with non-negative correlations
\be
\sum_{l=x,y,z}\langle j_l \otimes j_l\rangle_{\varrho_{AB}}\ge \sum_{l=x,y,z}\langle j_l\rangle_{\varrho}  \langle j_l\rangle_{\sigma} .\label{eq:corr1}
\ee
For such states, the entanglement condition
\be
\sum_{l=x,y,z}\langle j_l \otimes j_l\rangle_{\varrho_{AB}}\le \frac 1 2 F(\varrho,\sigma)-\frac 1 4,\label{eq:fident}
\ee
 is optimal, since it detects all entangled states $\varrho_{AB}$ within this set that can be detected based on knowing $\sum_{l=x,y,z}\langle j_l \otimes j_l\rangle_{\varrho_{AB}},$ $\varrho$ and $\sigma.$
 
 {\it Proof.} The inequality in \EQ{eq:corr1} is saturated for $\varrho_{AB}=\varrho\otimes\sigma.$ Due to Table~\ref{tab:max}(d), \EQ{eq:fident} is true for separable states and it can also be saturated by a separable state. $\qed$
 
 \subsection{B. Spin chains}
 
\begin{figure}[t!]
\includegraphics[width=0.45\columnwidth]{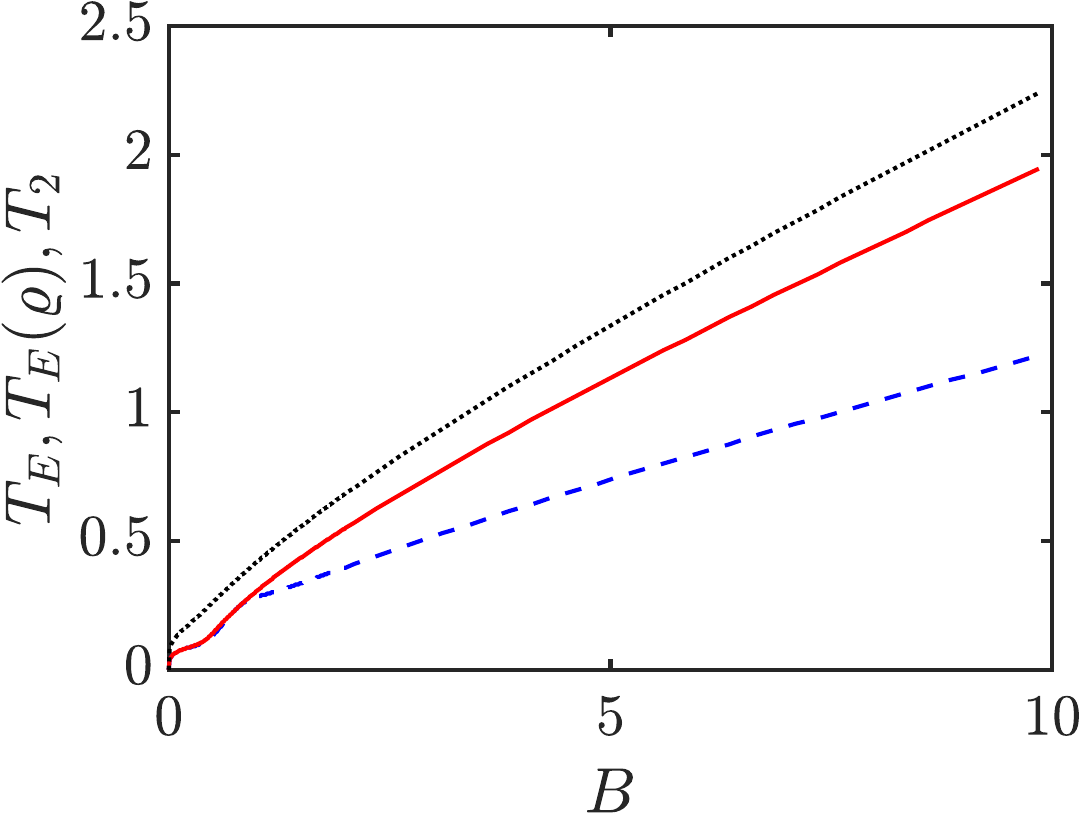}

\caption{Entanglement temperature as a function of the external field. Ferromagnetic Ising spin chain of $N=8$ spins in a transverse field.
(solid) Using \THM{thm:esepbound} to determine the temperature bound. (dashed) Bound without the knowledge of the single-particle states. (dotted) Entanglement temperature based on two-body nearest-neighbor entanglement.
}

 \label{fig:FQ_temperature_bound}
 \end{figure}

After bipartite systems, let us now consider spin chains. It has been mentioned in the introduction that the energy minimum for separable states can be used to detect entanglement in the thermal states of spin models \cite{Toth2005EntanglementWitnesses,Toth2006Detection,Guhne2006Energy,Guhne2005Multipartite,Brukner2004MacroscopicB,Dowling2004Energy,Wu2005Entanglement,Igloi2023Entanglement}. In this paper, we determine the energy bound knowing the single-particle state. This way we can improve the bound considerably, see \FIG{fig:FQ_temperature_bound}.
 
  \subsection{C. Systems on a complete graph}
 
 Apart from entanglement conditions for spin chains, we can also have entanglement criteria from Hamiltonians with interactions on a complete graph. The entanglement conditions for two-spin systems can be transformed to entanglement conditions for spin ensembles. For instance, \EQ{eq:entcond} leads to the criterion
\be
\ex{J_z^2}_{\varrho_N}\le N^2\ex{j_z^2}_{\overline\varrho}-\frac{N(N-1)}{4}{\mathcal F}_Q[\overline{\varrho},j_z].\label{eq:critfq}
\ee
If the criterion in \EQ{eq:critfq} is violated then the $N$-spin state $\varrho_N$ is entangled. Here, the average single-spin state is 
\be
\overline{\varrho}=\frac1{N}\sum_{n=1}^N\varrho^{(n)},
\ee
where $\varrho^{(n)}$ is the reduced state of the spin $n.$ We also used that due to the convexity of the quantum Fisher information
\be
{\mathcal F}_Q[\overline{\varrho},j_z]\le \frac{1}{N}\sum_{n=1}^N{\mathcal F}_Q[\varrho^{(n)},j_z]
\ee
holds. These criteria are related to the generalized spin squeezing inequalities, which are given in terms of first and second moments of collective observables in a system of $N$ spin-$j$ particles \cite{Sorensen2001Many-particle,Korbicz2005Spin,Toth2007Optimal,Toth2009Spin,Vitagliano2011Spin,Vitagliano2014Spin,Vitagliano2025sudsqueezingmany}. 

\section {SM4. Proof of \THM{thm:FQbound}}
\textlabel{app:thm:FQbound}{SM4}

For the complete-graph ground-state ensemble, the correlation estimator in Eq.~\eqref{eq:FQbound} can equivalently be written in terms of energy and the local field contribution:
\begin{align}
\mathcal F_Q^{(N)}[\varrho,h_1]
&=\frac8{N(N-1)J}\ex{H}_{\omega_N}\nonumber\\
&\quad+\frac8{(N-1)J}\vec B\ex{\vec g}_\varrho
+4\ex{h_1^2}_\varrho.\label{eq:FQbound_energy}
\end{align}
The finite-size error $\Delta=\mathcal F_Q[\varrho,h_1]-\mathcal F_Q^{(N)}[\varrho,h_1]$ satisfies
\begin{equation}
\Delta\le\frac1N[\lambda_{\max}(h_1)-\lambda_{\min}(h_1)]^2=O(N^{-1}).\label{eq:Delta}
\end{equation}
Permutation-invariant extensions are also connected to finite quantum de Finetti estimates \cite{Caves2002Unknown,Christandl2007One-and-a-Half} and monogamy relations \cite{Coffman2000Distributed,Osborne2006General}. The bounds proved here follow directly from the ground-state variational principle and QFI inequalities.

We restate the correlation bound in \EQ{eq:FQbound} as 
\be
{\mathcal F}_Q[\varrho,h_1]\ge{\mathcal F}_Q^{(N)}[\varrho,h_1]=4[\ex{h_1^2}_{\varrho}-C_N(\varrho)],\label{eq:FQbound2b}
\ee
where we define the quantity depending on $N$ and $\varrho$ as 
\be
C_N(\varrho)=\frac{1}{N(N-1)}\max_{\varrho_N\in \mathcal D_N(\varrho)} \left \langle \sum_{m\ne n} h_1^{(m)}h_1^{(n)}\right \rangle_{\varrho_N}.\label{eq:CN}
\ee


We now prove the inequality given in \EQ{eq:FQbound2b}.
Let $P_g$ denote the projector onto the ground-state subspace
of $H$. We consider the zero-temperature thermal state
\begin{equation}
\omega_N=\frac{P_g}{{\rm Tr}(P_g)}.\label{eq:omegaNPg}
\end{equation}
Since the Hamiltonian is permutationally invariant,
$\omega_N$ is also permutationally invariant, even if the
ground state is degenerate. We denote its single-particle
reduced state by $\varrho$.

For the ground-state energy and the energy minimum for
symmetric separable states we have
\be
E_g\le E_{\rm symsep}(\varrho).
\label{eq:EgEsymsep}
\ee
Using \EQS{eq:EgEsymsep} and \eqref{eq:maxhhFq}, we have
\be
\langle h_1^2\rangle_\varrho
-\frac{1}{N(N-1)}
\left\langle
\sum_{m\ne n}h_1^{(m)}h_1^{(n)}
\right\rangle_{\omega_N}
\le \frac14 {\mathcal F}_Q[\varrho,h_1],
\ee
which is equivalent to \EQ{eq:FQbound2b}.
Thus, we have proved the inequality in \EQ{eq:FQbound}.

Now we will find an upper bound on
${\mathcal F}_Q[\varrho,h_1]$.
Let us define the following $N$-particle operator
\begin{equation}
G=h_1^{(1)}-\frac{1}{N-1}\sum_{n=2}^{N}h_1^{(n)}.
\end{equation}
Consider the family of density matrices
\begin{equation}
\omega_N(\theta)=e^{-i\theta G}\omega_N e^{i\theta G}.
\end{equation}
Since the two terms in $G$ act on different subsystems,
the unitary factorizes as
\begin{equation}
e^{-i\theta G}
=e^{-i\theta h_1^{(1)}}\otimes
e^{\,i\theta\frac{1}{N-1}\sum_{n=2}^{N}h_1^{(n)}}.
\end{equation}
Thus, after tracing out particles $2,\ldots,N$,
the reduced state of the first particle is
\begin{equation}
\varrho(\theta)=e^{-i\theta h_1}\varrho e^{i\theta h_1}.
\end{equation}
Using the monotonicity of the quantum Fisher information
under the partial trace and the general inequality
${\mathcal F}_Q[\omega_N,G]\le 4\va{G}_{\omega_N}$,
we obtain
\begin{equation}
{\mathcal F}_Q[\varrho,h_1]
\le {\mathcal F}_Q[\omega_N,G]
\le 4\va{G}_{\omega_N}
=4\langle G^2\rangle_{\omega_N}.
\label{eq:FQGbound}
\end{equation}
The last equality follows from permutation invariance,
which implies $\langle G\rangle_{\omega_N}=0$.

For the expectation value of the quadratic term we obtain
\begin{equation}
\langle G^2\rangle_{\omega_N}
=\frac{N}{N-1}
[\langle h_1^2\rangle_{\varrho}-C_N(\varrho)].
\label{eq:G2}
\end{equation}
Here, $C_N(\varrho)$ is defined as in \EQ{eq:CN},
and the equality follows because $\omega_N$ minimizes
the energy among states with the same single-particle
marginal $\varrho$.

Using the inequalities given in \EQS{eq:FQGbound}
and \eqref{eq:G2}, it follows that
\begin{equation}
{\mathcal F}_Q[\varrho,h_1]
\le \frac{4N}{N-1}
[\langle h_1^2\rangle_{\varrho}-C_N(\varrho)].
\label{eq:FQbound2bb}
\end{equation}

The inequalities in \EQS{eq:FQbound2b}
and \eqref{eq:FQbound2bb} imply \EQ{eq:rendorelv},
where ${\mathcal F}_Q^{(N)}[\varrho,h_1]$ denotes
the right-hand side of \EQ{eq:FQbound},
as well as the right-hand side of \EQ{eq:FQbound2b}.
Denoting the difference between the left-hand and
right-hand sides of \EQ{eq:FQbound} by $\Delta$, we obtain
\begin{align}
\begin{split}
&0\le\Delta
\le\frac1N{\mathcal F}_Q[\varrho,h_1]
\le\frac4N\va{h_1}_\varrho\\
&\qquad\qquad\qquad
\le\frac1N
[\lambda_{\max}(h_1)-\lambda_{\min}(h_1)]^2,
\end{split}
\end{align}
which finishes our proof.
The bounds therefore remain valid for degenerate ground
states when the system is described by its zero-temperature
thermal state.

Note that the de Finetti theorem also shows that
the two-body reduced state approaches the set of separable
states as $N$ increases
\cite{Caves2002Unknown,Christandl2007One-and-a-Half}. 
In our setting, however, we additionally require the
separable state to have the prescribed single-particle
reduced state $\varrho$. $\qed$

\EQL{eq:rendorelv} implies that for $N\rightarrow\infty$ we get back the quantum Fisher information
 \be
\lim_{N\rightarrow\infty}{\mathcal F}_Q^{(N)}[\varrho,h_1] = {\mathcal F}_Q[\varrho,h_1],
\ee
while for $N=2$ and $d=2$ we have 
\be
{\mathcal F}_Q^{(2)}[\varrho,h_1] =4I_{\varrho}^{\rm WY}(h_1),
\ee
 see Table~\ref{tab:max}(c). 
 
Next, we prove the following inequality
\be
{\mathcal F}_Q^{(2)}[\varrho,h_1]\le{\mathcal F}_Q^{(3)}[\varrho,h_1]\le ... \le{\mathcal F}_Q[\varrho,h_1]. \label{eq:ineq}
\ee
This can be seen by writing ${\mathcal F}_Q^{(N)}$ as in \EQ{eq:FQbound2b}. If $\varrho_{N+1,{\rm opt}}$ is the optimal state for $N+1$ particles, then we can obtain an $N$-particle state that has the same average two-body correlations as
\be
\operatorname{Tr}_1(\varrho_{N+1,{\rm opt},{\rm PI}}),
\ee
where the permutation-invariant state is defined as
\be
\varrho_{N+1,{\rm opt},{\rm PI}}=\frac 1 {(N+1)!} \sum_k \Pi_k \varrho_{N+1,{\rm opt}} \Pi_k^\dagger,
\ee
and the summation is over all permutations. However, there can be other $N$-particle states with the same marginals $\varrho$ that give a larger value. Hence,
\be
C_N(\varrho)\ge C_{N+1}(\varrho)
\ee
holds, which ends the proof of \EQ{eq:ineq}. \qed

We will now show that ${\mathcal F}_Q^{(N)}[\varrho,h_1]$ is a generalized quantum Fisher information (see, e.g., \REF{Toth2013Extremal}). 
Let us denote the energy minimum for a ferromagnetic Ising spin system on a fully connected graph for $N$ particles as  
\be
E_N(\varrho)=\min_{\varrho_N\in{\mathcal D}_N(\varrho)}\ex{H}_{\varrho_N},\label{eq:EN}
\ee
which is convex in $\varrho.$ The right-hand side of \EQ{eq:FQbound} can be given as
\begin{align}
\begin{split}
{\mathcal F}^{(N)}_Q[\varrho,h_1]&=\frac 8 {N(N-1)J} E_N(\varrho)\\
&+\frac 8{N(N-1)J}N\vec B\ex{\vec g}_\varrho
+4\ex{h_1^2}_{\varrho}.\label{eq:FQbound2}
\end{split}
\end{align}
Hence, (i) ${\mathcal F}_Q^{(N)}[\varrho,h_1]$ is also convex in the state. The derivation leading to \EQ{eq:rendorelv} applies also to $\mathcal F_Q^{(N)}[\varrho,h_1]$ defined in \EQ{eq:FQbound2}. If $\varrho$ is pure then the minimizing state is $\varrho^{\otimes N},$ thus  (ii) ${\mathcal F}_Q^{(N)}[\varrho,h_1]$ given in \EQ{eq:FQbound2} equals four times the variance. Due to (i) and (ii), ${\mathcal F}_Q^{(N)}[\varrho,h_1]$ is a generalized quantum Fisher information \cite{Toth2013Extremal}.

\section{SM5. Ferromagnetic Heisenberg spin system} 
\textlabel{sec:ferro}{SM5}

\subsection{A. Ferromagnetic Heisenberg spin chain} 

Let us consider a ferromagnetic Heisenberg chain of spin-$1/2$ particles with the Hamiltonian given in \EQ{eq:H} with $L=3$ and $J_1=J_2=J_3=-J<0$ and $h_1=j_x, h_2=j_y, h_3=j_z.$ Applying the two-colorable construction of \THM{thm:The minimum for separable states} to alternating marginals, for even $N$ we have a tight lower bound for separable states
\begin{align}
\begin{split}
\min_{\varrho_{N}\in\mathrm{Sep}_N(\varrho,\sigma,\varrho,\sigma, ... )}  \langle H\rangle_{\varrho_{N}}&=-JN\left[\frac 1 2 F(\varrho,\sigma)- \frac 1 4\right]\\
&- N \vec B \frac{\ex{\vec\sigma}_{\varrho}+\ex{\vec\sigma}_{\sigma}}{2},
\end{split}
\end{align}
where $\mathrm{Sep}_N(\varrho_1,\varrho_2, ...)$ denotes the set of $N$-qubit separable states with the marginals $\varrho_n.$
Here, the Uhlmann--Jozsa fidelity is defined as \cite{Uhlmann1976TheTransitionProbability,Jozsa1994Fidelity}
\be
F(\varrho,\sigma)={\rm Tr}\left(\sqrt{\sqrt{\varrho} \sigma \sqrt{\varrho}}\right)^2.
\ee

\subsection{B. Ferromagnetic Heisenberg system on a bipartite graph} 

Next, we demonstrate how the fidelity can be extracted from correlation measurements. To this end, we consider a spin system defined on a bipartite graph rather than a conventional spin chain. The $N$ particles (or qubits) are partitioned into two disjoint groups, which we label $A$ and $B$. There are no intra-group interactions. Instead, every spin in group $A$ couples to every spin in group $B$ via the ferromagnetic Heisenberg interaction. The corresponding $N$-qubit ferromagnetic Hamiltonian is
\be
H=- J \sum_{l=x,y,z}\sum_{n=1}^{N_1}\sum_{n'=N_1+1}^N j_l^{(n)}j_l^{(n')}-\vec B_1\vec G_1-\vec B_2\vec G_2,\label{eq:xxyyzz}
\ee
where 
\begin{equation}
\vec G_1=\left(\sum_{n=1}^{N_1} g_1^{(n)}, \sum_{n=1}^{N_1} g_2^{(n)}, \sum_{n=1}^{N_1} g_3^{(n)}, ... \right),
\end{equation}
and
\begin{equation}
\vec G_2=\left(\sum_{n=N_1+1}^{N} g_1^{(n)}, \sum_{n=N_1+1}^{N} g_2^{(n)}, \sum_{n=N_1+1}^{N} g_3^{(n)}, ... \right),
\end{equation}
and $\vec g=\{g_l\}_{l=1}^{d^2-1}$ are the $SU(d)$ generators.
Here particles in the first group, that is particles 1, 2, ..., $N_1$ have the reduced state $\varrho.$ Particles in the other group, i.e., particles $N_1+1, N_1+2,$ ..., $N$ have the reduced state $\sigma.$ $\vec B_1$ and $\vec G_1$ correspond to the first particle group, while $\vec B_2$ and $\vec G_2$ correspond to the second one. 

Let $P_g$ denote the projector onto the ground-state
subspace of the Hamiltonian. We consider the
zero-temperature thermal state given in \EQ{eq:omegaNPg}.
The ground-state energy yields an upper bound on the
quantum fidelity of two qubits as
\begin{align}
\begin{split}
F(\varrho,\sigma)&\le\frac1 2-\frac2{N_1N_2J} 
 \ex{H}_{\omega_N}\\
&-\frac 2{N_1N_2 J}(N_1 \vec B_1 \ex{\vec \sigma}_\varrho+N_2  \vec B_2 \ex{\vec \sigma}_\sigma),
\label{eq:ineqFid}
\end{split}
\end{align}
where $N_2=N-N_1$ and $\vec \sigma=(\sigma_x,\sigma_y,\sigma_z).$

The Hamiltonian in \EQ{eq:xxyyzz} is invariant under permutations of particles within each of the two groups. Since the Hamiltonian is invariant under permutations within each group, so is $\omega_N$, even if the ground state is degenerate. Consequently, its two-particle reduced state $\varrho_{AB}$, consisting of one particle from each group, admits permutation-invariant extensions to $N_1$ particles in group $A$ and $N_2$ particles in group $B$.

Let $M=\min(N_1,N_2)$. By the finite quantum de Finetti theorem for permutation-invariant extensions~\cite{Christandl2007One-and-a-Half}, there exists a separable two-particle state $\tilde{\varrho}_{AB}$ whose trace distance from $\varrho_{AB}$ tends to zero as $M$ increases. Consequently, the single-particle marginals $\tilde{\varrho}$ and $\tilde{\sigma}$ of $\tilde{\varrho}_{AB}$ converge to $\varrho$ and $\sigma$, respectively.

The two-qubit correlation identity used for the Heisenberg-chain minimum is
\begin{equation}
\max_{\varrho_{AB}\in\mathrm{Sep}_2(\varrho,\sigma)}\sum_{l=x,y,z}\ex{j_l\otimes j_l}_{\varrho_{AB}}=\frac12F(\varrho,\sigma)-\frac14.\label{eq:fid}
\end{equation}
For two qubits, \EQ{eq:fid} is equivalent to
\begin{equation}
\max_{\eta_{AB}\in{\rm Sep}_2(\varrho,\sigma)}
\langle F_{AB}\rangle_{\eta_{AB}}
=F(\varrho,\sigma).
\label{eq:ineqFid2}
\end{equation}
Here $F_{AB}$ denotes the flip operator. Applying this relation to the separable state $\tilde{\varrho}_{AB}$ with marginals $\tilde{\varrho}$ and $\tilde{\sigma}$ gives
\begin{equation}
\langle F_{AB}\rangle_{\tilde{\varrho}_{AB}}
\le F(\tilde{\varrho},\tilde{\sigma}).
\label{eq:ineqfab}
\end{equation}
Since $\tilde{\varrho}_{AB}$ converges to $\varrho_{AB}$ in trace distance, while $\tilde{\varrho}\to\varrho$ and $\tilde{\sigma}\to\sigma$, the continuity of operator expectation values and of the fidelity, together with \EQ{eq:ineqfab}, implies that for every $\varepsilon>0$, for sufficiently large $M$,
\begin{equation}
\langle F_{AB}\rangle_{\varrho_{AB}}
\le F(\varrho,\sigma)+\varepsilon.
\end{equation}
On the other hand, \EQ{eq:ineqFid} implies
\begin{equation}
F(\varrho,\sigma)
\le \langle F_{AB}\rangle_{\varrho_{AB}}.
\end{equation}
Combining the two relations, we obtain
\begin{equation}
\lim_{M\rightarrow\infty}
\left[
\langle F_{AB}\rangle_{\varrho_{AB}}
-F(\varrho,\sigma)
\right]=0.
\end{equation}
Hence, the error $\Delta$ of the fidelity estimate in \EQ{eq:ineqFid} tends to zero as $M\rightarrow\infty$.

\begin{figure}[t!]
\includegraphics[width=0.45\columnwidth]{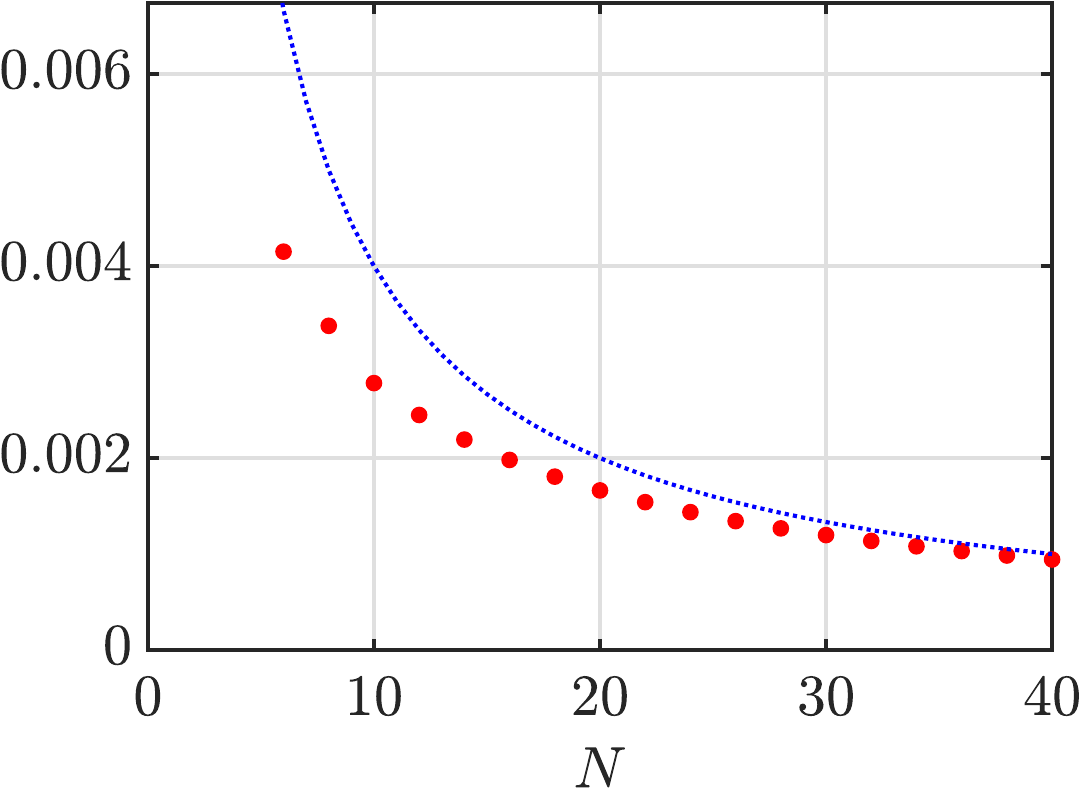}\hskip0.4cm\includegraphics[width=0.45\columnwidth]{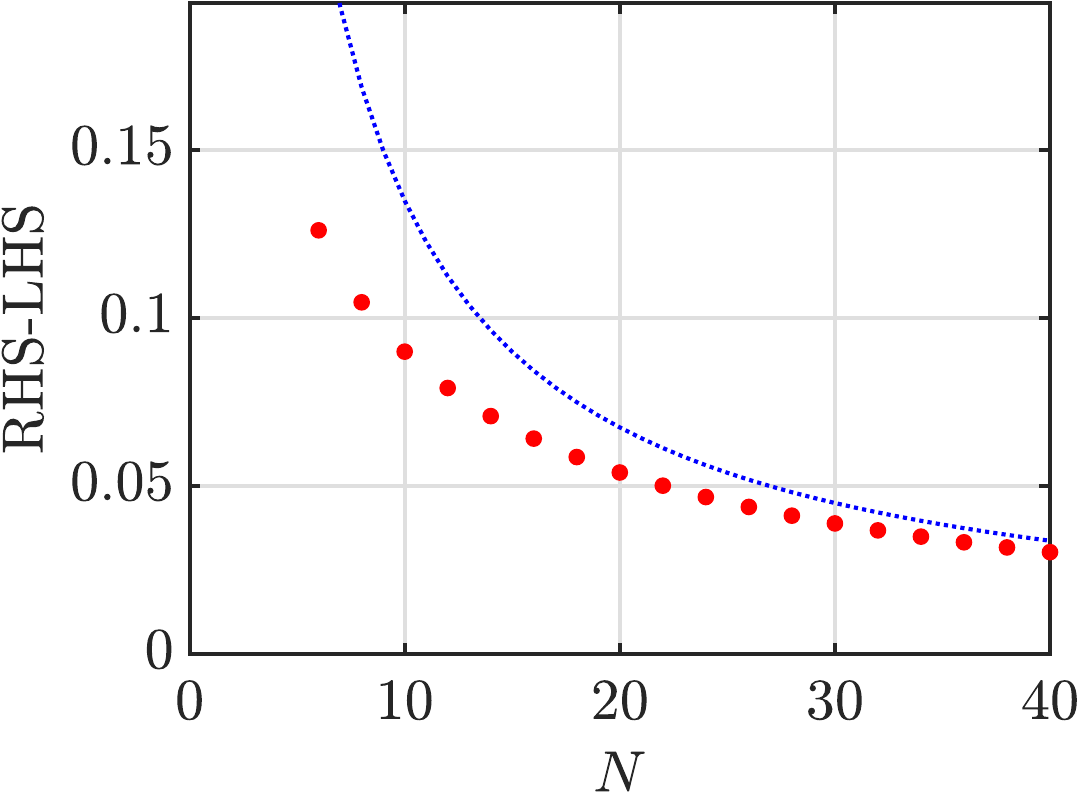}

\hspace*{0.55cm}(a)\hskip3.9cm(b)
\caption{Estimating the two-qubit fidelity from correlation measurements using \EQ{eq:ineqFid} for $N=6,8,10,...,40.$ (a) Curves for the states given in \EQ{eq:rho1}. (disks) Difference between the right-hand side and the left-hand side of \EQ{eq:ineqFid} as a function of $N$ for a given pair of density matrices. (dotted) $0.04/N$, shown as a guide to the eye.
(b) Curves for the states given in \EQ{eq:rho2}. (disks) Difference between the right-hand side and the left-hand side of \EQ{eq:ineqFid} as a function of $N$ for a given pair of density matrices. (dotted) $1.35/N$, shown as a guide to the eye.
} \label{fig:FQ_ground2}
\end{figure}

The numerical test of the bound given in \EQ{eq:ineqFid}  can be seen in \FIG{fig:FQ_ground2}(a). We consider  the case $N_1=N_2=N/2.$ The density matrices for the first calculation are
\be
\varrho=
\left(
\begin{array}{cc}
  0.9500 &   0.1750  \\
  0.1750 &   0.0500  \\
\end{array}
\right)
,\quad 
\sigma= 
\left(
\begin{array}{cc}
  0.8250 &   0.3500  \\
  0.3500 &   0.1750  \\
\end{array}
\right).\label{eq:rho1}
\ee
The fidelity between the two density matrices is $0.9534.$
See \FIG{fig:FQ_ground2}(a).

We carried out another test with the density matrices
\begin{align}
\begin{split}
\varrho&=
\left(
\begin{array}{cc}
  0.2337 &   -0.2663 + 0.2663i  \\
  -0.2663 - 0.2663i &   0.7663  \\
\end{array}
\right)
,\\
\quad 
\sigma&=
\left(
\begin{array}{cc}
  0.7663 &   0.2663 - 0.2663i  \\
  0.2663 + 0.2663i &   0.2337  \\
\end{array}
\right).\label{eq:rho2}
\end{split}
\end{align}
The fidelity between the two density matrices is $0.1490.$
 The results can be seen in \FIG{fig:FQ_ground2}(b).

\section{SM6. Antiferromagnetic Ising spin systems with a symmetric ground state}
\textlabel{sec:antiferro}{SM6}

In this section, we consider a minimization problem that is important in entanglement theory and the theory of spin squeezing  \cite{Sorensen2001Entanglement}. 

Let us consider the ground state of
\be
H=J_x^2-\lambda J_z
\ee
for $N$ spin-$1/2$ particles, where the single-particle density matrix is $\varrho,$ $\lambda$ is a constant, and the collective angular momentum components are defined as
\be
J_l=\frac1 2 \sum_{n=1}^{N}\sigma_l^{(n)}
\ee
for $l=x,y,z.$ The minimum for symmetric separable states is  
\begin{align}
E_{\rm symsep}(\varrho)&=\frac N 4+N(N-1)\ex{j_x}^2_{\varrho}-\lambda N\ex{j_z}_{\varrho}\nonumber\\
&=\frac N 4+\frac{N-1}N \ex{J_x}^2-\lambda\ex{J_z}. 
\end{align}
Note that the state giving the energy minimum is symmetric. Hence, we can use that  

\be
\min_{\varrho_{AB}\in\mathrm{SymSep}_2(\varrho)}\ex{h\otimes h}_{\varrho_{AB}}=\ex{h}_{\varrho}^2,\label{eq:minS2prime}
\ee
where $\mathrm{SymSep}_2(\varrho)$ is the set of bipartite symmetric separable states with marginal $\varrho,$ see Table~\ref{tab:min}(b). 

Let us see the general case with any positive integer $L$ and arbitrary $h_l,$ which includes, for example, the XX model
\be
H= \sum_{l=1}^{L} \left(\sum_{n=1}^{N}h_l^{(n)}\right)^2 -\lambda J_z.
\ee
For this case we have 
\be
\min_{\varrho_{AB}\in\mathrm{SymSep}_2(\varrho)}\sum_{l=1}^L\ex{h_l\otimes h_l}_{\varrho_{AB}}\ge\sum_{l=1}^L\ex{h_l}_{\varrho}^2,\label{eq:L123}
\ee
where for $L=1,2$ there is an equality,  see Table~\ref{tab:min}(b-c). 
Then, for the minimum for symmetric separable states 
\begin{equation}
\begin{aligned}
E_{\rm symsep}(\varrho)&\ge N\sum_{l=1}^L\ex{h_l^2}_{\varrho}\\
&\quad+N(N-1)\sum_{l=1}^L\ex{h_l}^2_{\varrho}
-\lambda N\ex{j_z}_{\varrho}.
\end{aligned}\label{eq:sepbound}
\end{equation}
holds, while for $L=1,2$ there is an equality in \EQ{eq:sepbound}.

\section{SM7. Multipartite entanglement}
\textlabel{sec:multi}{SM7}

In this section, we first reduce general $k$-producible states to mixtures of states separable across contiguous-block partitions. We then derive energy bounds for the more restricted class associated with a fixed partition into equal-size contiguous blocks, and also obtain lower bounds valid for arbitrary states.

Pure $k$-producible states are given as 
\be
|\Psi_{k-{\rm prod}}\rangle=\bigotimes_{m=1}^M \ket{\Psi^{(m)}},
\ee
where $\ket{\Psi^{(m)}}$ are states of at most $k$ particles. Mixed $k$-producible states are mixtures of pure $k$-producible states. States that are not $k$-producible are at least $(k+1)$-particle entangled \cite{Guhne2005Multipartite,Guhne2006Energy}.

Next, we show that it is sufficient to consider blocks of consecutive spins.

\DEFTHM{thm:cblocks}For a nearest-neighbor spin chain, every $k$-producible
state can be replaced by a mixture of states separable
with respect to partitions into contiguous blocks of
at most $k$ particles, preserving the energy and all
single-particle marginals.

\textit{Proof.}
Consider a pure $k$-producible state
$|\Psi\rangle=\bigotimes_{B\in\mathcal P}|\psi_B\rangle$,
where $|B|\leq k$.
Here, $\mathcal P$ denotes a partition of the $N$ particles into disjoint blocks, e.g., $\mathcal P=\big\{\{1,3\},\{2,5\},\{4,6\}\big\}.$
Split each block $B$ into its maximal connected
components $C_i$ along the chain, and replace
$|\psi_B\rangle\langle\psi_B|$ by
\begin{equation}
\bigotimes_i
\operatorname{Tr}_{B\setminus C_i}
(|\psi_B\rangle\langle\psi_B|).
\end{equation}
Every nearest-neighbor pair originally belonging
to the same block remains in the same component,
so its reduced density matrix is unchanged.
Pairs belonging to different blocks retain their
product marginals. Thus, all nearest-neighbor
correlations and single-particle marginals are preserved.

Decomposing the resulting block states into pure
states gives a  mixture of contiguous-block
$k$-producible states. Applying the construction
to each component of a mixed state completes
the proof. $\qed$

We now restrict to the fixed partition $\mathcal P_k$ of Appendix~\ref{Appendix C: Multipartite entanglement}, with $N$ divisible by $k$ and all blocks containing exactly $k$ consecutive particles. Its separable states have the form
\be
\sum_i p_i \ketbra{\Psi_i^{(1)}}\otimes\ketbra{\Psi_i^{(2)}}\otimes...\otimes\ketbra{\Psi_i^{(M)}},\label{eq:twoprod}
\ee
with all single-particle marginals equal to $\varrho$; these states form $\mathcal E^{(k)}_N(\varrho)$. Their minimum $E_{\mathcal P_k}(\varrho)$ is no smaller than the minimum over all $k$-producible states. A lower bound on $E_{\mathcal P_k}(\varrho)$ therefore need not bound the unrestricted minimum. The preceding contiguous-block reduction allows variable block sizes and partitions and does not remove this restriction.

We first derive a lower bound for this fixed partition.

\DEFTHM{thm:The minimum for k-producible states}For a periodic $N$-particle spin chain, the minimum energy over states separable across $\mathcal P_k$ with single-particle marginal $\varrho$ obeys 
\begin{align}
\begin{split}
E_{\mathcal P_k}(\varrho)&:=\min_{\varrho_N\in\mathcal E^{(k)}_N(\varrho)}\ex{H}_{\varrho_N}\\
&\ge\frac N k\min_{\varrho_{k\text{-part}}\in\mathcal D_k(\varrho)}\ex{H_{k\text{-part}}}_{\varrho_{k\text{-part}}}\\
&+\frac N k\min_{\varrho_{AB}\in\mathrm{Sep}_2(\varrho)}\ex{H_{AB}}_{\varrho_{AB}},\label{eq:Hbound2}
\end{split}
\end{align}
where $H_{nn'}$ is given in \EQ{eq:Hmn}, $N_p=N,$ $N$ is divisible by $k,$  $N>k,$ and the $k$-particle Hamiltonian is defined as in \EQ{eq:Hhpart}.

{\it Proof.} The proof is analogous to that of \EQ{eq:Hbound} in \THM{thm:The minimum for separable states}, i.e., the minimum of a sum is at least as large as the sum of the individual minima.

Note that since both the Hamiltonian and the set of states considered are invariant under cyclic translations by $k$ sites, the minimizing state can be chosen invariant under such block translations. Hence, all $k$-particle blocks can be assumed to have the same reduced state $\varrho_{k\text{-part}}$.
$\qed$

In \EQ{eq:Hbound2}, the first term is the contribution of the $k$-particle units; the second term is an expectation value of the interaction term between these units minimized for a separable state with given marginals.

As an application, for a ferromagnetic spin chain with a single nearest-neighbor interaction term discussed in \THM{thm:esepbound_kprod}, we obtain a lower bound for states separable across $\mathcal P_k$. Let us first consider two neighboring blocks. Let us define
\be
h_L=h_1\otimes\openone^{\otimes(k-1)},\qquad h_R=\openone^{\otimes(k-1)}\otimes h_1.
\ee
Then the coupling between two neighboring blocks is
\be
-J\,h_R\otimes h_L.
\ee
For a separable two-block state
\be
\omega_{AB}=\sum_i p_i|\Psi_i\rangle\langle\Psi_i|\otimes|\Phi_i\rangle\langle\Phi_i|,
\ee
with both block marginals equal to $\varrho_{k\text{-part}},$

\be
\langle h_R\otimes h_L\rangle_{\omega_{AB}}=\sum_i p_i\langle h_R\rangle_{\Psi_i}\langle h_L\rangle_{\Phi_i}.
\ee
Using the well-known inequality
\be
xy\le \frac{x^2+y^2}{2},
\ee
we obtain
\begin{align}
\begin{split}
&\langle h_R\otimes h_L\rangle_{\omega_{AB}}\le\frac12\sum_i p_i\left(\langle h_R\rangle_{\Psi_i}^2+\langle h_L\rangle_{\Phi_i}^2\right)\\
&\qquad\le \frac12\ex{h_R^2+h_L^2}_{\varrho_{k\text{-part}}}\\
&\qquad-\frac1 8\mathcal F_Q[\varrho_{k\text{-part}},h_R]-\frac1 8\mathcal F_Q[\varrho_{k\text{-part}},h_L].
\end{split}
\end{align}
Since every one-particle marginal of the block is $\varrho,$
\be
\langle h_L^2\rangle_{\varrho_{k\text{-part}}}=\langle h_R^2\rangle_{\varrho_{k\text{-part}}}=\langle h_1^2\rangle_\varrho.
\ee
Therefore,
\begin{align}
\begin{split}
&\max_{\omega_{AB}\in\mathrm{Sep}_2(\varrho_{k\text{-part}})}\langle h_R\otimes h_L\rangle_{\omega_{AB}}\le\langle h_1^2\rangle_\varrho\\
&\qquad\qquad-\frac18\left\{\mathcal F_Q[\varrho_{k\text{-part}},h_L]+\mathcal F_Q[\varrho_{k\text{-part}},h_R]\right\}.
\end{split}
\end{align}
For a fixed block state $\varrho_{k-\rm part}$, the lower bound
\begin{align}
&E_{\mathcal P_k}(\varrho_{k-\rm part})\ge\frac N k \ex{H_{k-\rm part}}_{\varrho_{k-\rm part}} \nonumber\\
&\quad\quad\quad-\frac N k J\left(\ex{h_1^2}_\varrho-\frac1 4 \overline{\mathcal F}_Q(\varrho_{k-\rm part})\right)- \frac {N}k\vec B\ex{\vec g}_\varrho\label{eq:bound2k}
\end{align}
holds, where the quantum Fisher information averaged over the two boundary operators is
\be
\overline{\mathcal F}_Q(\varrho_{k-\rm part})= \frac1 2\left( {\mathcal F}_Q[\varrho_{k-\rm part},h_L] + {\mathcal F}_Q[\varrho_{k-\rm part},h_R]\right),
\ee
all $k$-particle units have the reduced state $\varrho_{k-\rm part},$ and based on \EQ{eq:Hhpart} we have
\begin{align}
\begin{split}
H_{k\text{-part}}&=-J\sum_{n=1}^{k-1}h_1^{(n)}h_1^{(n+1)}-\vec B\sum_{n=1}^k\vec g^{(n)}\\
&+\frac{1} 2\vec B(\vec g^{(1)}+\vec g^{(k)}).
\end{split}
\end{align}
With this choice, the total Hamiltonian in \EQ{eq:H} can be written as 
\begin{align}
\begin{split}
H&=\sum_{n=0}^{N/k-1} H_{k\text{-part}}^{(nk+1,nk+2,nk+3,...,(n+1)k)}\\
&+\sum_{n=0}^{N/k-2}  H_{nk+k,nk+k+1}+H_{N,1},
\end{split}
\end{align}
where the superscripts indicate the spins on which the Hamiltonian acts, and $H_{nn'}$ is again given in \EQ{eq:Hmn}. Now the bound is a function of the $k$-particle quantum state. This bound can be reformulated as 
\begin{align}
&E_{\mathcal P_k}(\varrho)\ge\min_{\varrho_{k-\rm part}\in\mathcal D_k(\varrho)} \bigg[\frac N k \ex{H_{k-\rm part}}_{\varrho_{k-\rm part}} \nonumber\\
&\quad\quad\quad-\frac N k J\left(\ex{h_1^2}_\varrho-\frac1 4 \overline{\mathcal F}_Q(\varrho_{k-\rm part})\right)- \frac {N}k\vec B\ex{\vec g}_\varrho\bigg],\label{eq:bound2kb}
\end{align}
where we minimize over all $k$-particle states having $\varrho$ marginals, and  $J_1=-J<0.$

If we compare  \EQ{eq:bound2k} and \EQ{eq:bound2kb} with each other, we can see that in   \EQ{eq:bound2k}  the $k$-particle state is constrained, while in \EQ{eq:bound2kb} the single-particle state is constrained.

We can also formulate another bound with the quantum Fisher information given in \EQ{eq:bound2} in \THM{thm:esepbound_kprod}.

{\it Proof of \THM{thm:esepbound_kprod}.}  The right-hand side of \EQ{eq:bound2} is not larger than that of \EQ{eq:bound2kb}, since  due to the general properties of the quantum Fisher information, we have ${\mathcal F}_Q[\varrho_{k-\rm part}, h_R]\ge {\mathcal F}_Q[\varrho, h_1]$ and ${\mathcal F}_Q[\varrho_{k-\rm part}, h_L]\ge {\mathcal F}_Q[\varrho, h_1],$ hence $\overline{{\mathcal F}}_Q(\varrho_{k-\rm part})\ge {\mathcal F}_Q[\varrho, h_1].$ $\qed$
 
The preceding bounds concern states separable across $\mathcal P_k$. A lower bound valid for arbitrary states with the same single-particle marginals is obtained by allowing an arbitrary two-particle state at each block boundary:
\begin{align}
E_{\min}(\varrho)&\ge \frac N k\min_{\tau_k\in\mathcal D_k(\varrho)}
\ex{H_{k\text{-part}}}_{\tau_k}\nonumber\\
&\quad+\frac N k\min_{\tau_{AB}\in\mathcal D_2(\varrho)}
\ex{H_{AB}}_{\tau_{AB}}.\label{eq:Hbound2cc}
\end{align}
Here $H_{AB}$ is the bond Hamiltonian in \EQ{eq:Hmn} with $N_p=N$.
Comparing to \EQ{eq:bound2}, one can see that now the minimization in the second term is over physical states rather than over separable states. For a ferromagnetic $N$-qubit spin chain with a single nearest-neighbor interaction term discussed in \THM{thm:esepbound_kprod} we have 
\begin{align}
\begin{split}
E_{\min}(\varrho)&\ge \frac N k \min_{\varrho_{k\text{-part}}\in \mathcal D_k(\varrho)}\ex{H_{k\text{-part}}}_{\varrho_{k\text{-part}}}\\
&-\frac N k J\left[\ex{h^2}_\varrho-I^{\rm WY}_{\varrho}(h)\right]-\frac {N}k\vec B\ex{\vec g}_\varrho.\label{eq:Hbound2db}
\end{split}
\end{align}

An alternative lower bound can be obtained as follows.

\DEFTHM{thm:lower_bound_on_the_energy2}For an $N$-qudit spin chain with periodic boundary conditions, the minimal energy of the Hamiltonian  given in \EQ{eq:H} for given marginals can be bounded from below as
\begin{align}
\begin{split}
\min_{\varrho_N\in \mathcal D_N(\varrho)}\ex{H}_{\varrho_N}
&\ge \frac N {k-1} \min_{\varrho_{k\text{-part}}\in \mathcal D_k(\varrho)}\ex{H_{k\text{-part}}'}_{\varrho_{k\text{-part}}}\\
&=:E_{\rm LB}^{(k)}(\varrho),
\end{split}
\end{align}
where $2\leq k\leq N,$ $\varrho_{k\text{-part}}$ is a $k$-particle quantum state, and $H_{k\text{-part}}'$ is 
\begin{align}
\begin{split}
H_{k\text{-part}}'&=\sum_{n=1}^{k-1} \sum_{l=1}^L J_l h_l^{(n)}h_l^{(n+1)}-\vec B \sum_{n=2}^{k-1}\vec g^{(n)}\label{eq:Hk}\\
&-\frac{1}{2}\vec B(\vec g^{(1)}+\vec g^{(k)}).
\end{split}
\end{align}
Note that $E_{\rm LB}^{(2)}(\varrho)\equiv E_{\rm LB}(\varrho).$ The bound $E_{\rm LB}^{(k)}$ for $k\ge2$ could give an even better lower bound.

{\it Proof.} Consider all $N$ possible cyclic translations of a $k$-site block, with $2\leq k\leq N.$ Each nearest-neighbor interaction appears in exactly $k-1$ of these blocks. With the corresponding field terms distributed consistently,  
\be
H=\frac1{k-1}\sum_{n=1}^NH'_{k\text{-part},n,n+1,\ldots,n+k-1},
\ee
where the summation is over the Hamiltonians acting on the various $k$-particle groups. 
$\qed$

We present results for various $k$ in \FIG{fig:Ising_chain_ground_maintext}, where we set $B=0$ for simplicity. The lower bound for the energy per particle with the Wigner--Yanase skew information is given in \EQ{eq:Hbound2db}.

This approach can easily be extended to the XY model or the Heisenberg chain in a transverse field \cite{takahashi1999thermodynamics}. In these cases, instead of the quantum Fisher information, numerical calculations based on semidefinite programming would be necessary for the two-qubit boundary term. These calculations would be easy to carry out. The optimization over the $k$-particle blocks must be performed separately. The lower bound with the Wigner--Yanase skew information will have several terms with Wigner--Yanase skew information \cite{Toth2023QuantumWasserstein}.


\end{document}